\documentclass{article}

\usepackage{microtype}
\usepackage{graphicx}
\usepackage{booktabs} 

\usepackage{listings}

\usepackage{makecell}
\usepackage{multirow}
\usepackage{caption}
\usepackage{subcaption}
\usepackage{xcolor}

\usepackage[hyphens]{url}
\usepackage{hyperref}
\usepackage[hyphenbreaks]{breakurl}

\definecolor{mygreen}{rgb}{0,0.6,0}
\definecolor{mygray}{rgb}{0.5,0.5,0.5}
\definecolor{mymauve}{rgb}{0.58,0,0.82}

\usepackage[accepted]{mlsys2022}
\usepackage{comment}

\def\SysName{\textit{OneFlow}}
\newcommand{\sys}{\SysName}

\mlsystitlerunning{OneFlow: Redesign the Distributed Deep Learning Framework from Scratch}

\newcommand{\cwu}[1]{{\textcolor{red}{CWu: #1}}}

\begin{document}

\twocolumn[
\mlsystitle{\Large \bf \SysName{}: Redesign the Distributed Deep Learning Framework from Scratch}



\mlsyssetsymbol{equal}{*}

\begin{mlsysauthorlist}
\mlsysauthor{Jinhui Yuan}{oneflow}
\mlsysauthor{Xinqi Li}{oneflow}
\mlsysauthor{Cheng Cheng}{oneflow}
\mlsysauthor{Juncheng Liu}{oneflow}
\mlsysauthor{Ran Guo}{oneflow}
\mlsysauthor{Shenghang Cai}{oneflow}
\mlsysauthor{Chi Yao}{oneflow}
\mlsysauthor{Fei Yang}{zhejiang}
\mlsysauthor{Xiaodong Yi}{hku}
\mlsysauthor{Chuan Wu}{hku}
\mlsysauthor{Haoran Zhang}{upenn}
\mlsysauthor{Jie Zhao}{state}
\end{mlsysauthorlist}

\mlsysaffiliation{oneflow}{OneFlow Research.}
\mlsysaffiliation{zhejiang}{Zhejiang Laboratory, yangf@zhejianglab.com.}
\mlsysaffiliation{hku}{The University of Hong Kong, cwu@cs.hku.hk.}
\mlsysaffiliation{upenn}{University of Pennsylvania, haorz@seas.upenn.edu.}
\mlsysaffiliation{state}{State Key Laboratory of Mathematical Engineering and Advanced Computing, zjbc2005@163.com}

\mlsyscorrespondingauthor{Jinhui Yuan}{yuanjinhui@oneflow.org}

\mlsyskeywords{Machine Learning, MLSys}

\vskip 0.3in

\begin{abstract}
Deep learning frameworks such as TensorFlow and PyTorch provide a productive interface for expressing and training a deep neural network (DNN) model on a single device or using data parallelism. Still, they may not be flexible or efficient enough in training emerging large models on distributed devices, which require more sophisticated parallelism beyond data parallelism. Plugins or wrappers have been developed to strengthen these frameworks for model 
or pipeline parallelism, but they 
complicate the usage and implementation of distributed deep learning. 
Aiming at a simple, neat redesign of distributed deep learning frameworks for various parallelism paradigms, we present \SysName{}, a novel distributed training framework based on an \textit{SBP} (\textit{split}, \textit{broadcast} and \textit{partial-value}) abstraction and the actor model. \textit{SBP} enables much easier programming of data parallelism and model parallelism than existing frameworks, and the actor model provides a succinct runtime mechanism to manage the complex dependencies imposed by resource constraints, data movement and computation in distributed deep learning. We demonstrate the general applicability and efficiency of \SysName{} for training various large DNN models with case studies and extensive experiments. The results show that \SysName{} outperforms many well-known customized libraries built on top of the state-of-the-art frameworks. The code of \SysName{} is available at: \url{https://github.com/Oneflow-Inc/oneflow}.
\end{abstract}
]



\printAffiliationsAndNotice{}  


\vspace{-0.4cm}
\section{Introduction}
\vspace{-0.2cm}

Deep learning (DL) models have become increasingly complicated and large~\cite{devlin-etal-2019-bert, brown-gpt-2020, fedus-switch-2021, kaplan-scaling-2020}. 
Severe challenges arise for existing DL frameworks such as TensorFlow \cite{abadi-2016-tensorflow} and PyTorch \cite{paszke-2019-pytorch} for training large-scale DL models, 
which were designed in the early days without initially foreseeing the emerging requirements, e.g., model/pipeline parallelism of large models~\cite{brown-gpt-2020,huang-2019-nips,wang-2019-eurosys}.




Depending on the structure of neural networks (NN) and hardware configuration, various parallelism schemes 
find their best usage~\cite{tal-ddl-2019}. 
Data parallelism is especially suitable for DL models with a relatively small set of parameters (usually less than tens of millions of parameters), where near-linear speed-up can be achieved once 
back propagation maximally overlaps with gradient/parameter communication~\cite{jeaugey2017nccl,hashemi19tictac,peng-2019-sosp, jiang-osdi-2020}. 
Model parallelism and pipeline parallelism are for models with a more significant number of parameters, which probably cannot fit into a single device or the communication cost is too high for data parallelism. Stanza \cite{wu2018stanza} and DLPlacer \cite{pal2019optimizing} adopt data parallelism for training the convolutional layers and model parallelism for other layers in convolutional neural network (CNN) models. OptCNN \cite{jia-2018-icml} parallelizes CNN model training by splitting operations along batch and channel dimensions on homogeneous devices. Tofu \cite{wang-2019-eurosys} utilizes a partition-n-reduce method to split a single operation into sub-operations and deploy partitions on multiple GPUs. 
FlexFlow \cite{jia-2019-sysml} 
searches the SOAP (sample, operation, attribute, parameter) space to exploit parallelism within and across operations.


In the best case, a distributed DL framework should be able to automatically generate the physical execution plan for any chosen parallelism scheme, minimizing manual programming efforts of users. 
Then a more advanced requirement 
is that the framework should be able to find the most appropriate parallelism strategy for any combination of NN structure and hardware configuration~\cite{shazeer-2018-nips}. However, existing DL frameworks cannot even accomplish the first goal, 
i.e., flexibly supporting various parallelism strategies. This is the exact problem we aim to address in this paper, with a novel redesign of distributed training framework.  

Some emerging open-source projects 
develop dedicated systems or customized libraries for better support of model or pipeline parallelism. 
For example, HugeCTR~\cite{Oldridge2020MerlinAG} enables model parallelism 
for large-scale click-through rate estimation. Megatron-LMs~\cite{mohanmmad-megatron-2020,narayanan-megatron-2021} and DeepSpeed~\cite{deepspeed, rajbhandari-zeroinfinity-2021,rajbhandari-zero-2020} support model parallelism 
for pre-training large NLP models. InsightFace~\cite{insightface} trains large-scale face recognition models with model parallelism. 
However, these systems are customized for specific applications, and cannot 
be assembled together to constitute a general solution due to compatibility issues.

Wrappers or plugins have also been proposed to enhance 
some mainstream DL frameworks (e.g., TensorFlow, PyTorch) for better support of more complex parallelism schemes. Mesh-TensorFlow \cite{shazeer-2018-nips} and GShard \cite{dmitry-2020-gshard} provide APIs for developers to express a wide range of parallel computation patterns of DNNs on top of TensorFlow. GPipe \cite{huang-2019-nips} and PipeDream \cite{narayanan-2019-sosp} use pipelining across distributed devices to address the limited memory capacity on each device for training large DNNs on TensorFlow and PyTorch respectively. 
FairScale~\cite{fairscale} integrates techniques from Megatron-LM and DeepSpeed to enable PyTorch with model parallelism and pipeline parallelism. 
Since the existing training frameworks were initially designed without forseeing such complicated parallelism, 
incremental improvements over the frameworks often yield non-negligible system overhead and require substantial engineering efforts from users. 







What would a generic design and efficient implementation of distributed DL frameworks be if we could know the rapidly evolving large AI models and 
demand for various parallelism schemes in advance? Could the system be simpler and neater? In this paper, we explore such possibilities and present \SysName{}, a novel DNN training framework built from scratch. 
\SysName{} includes a holistic design from the compiler to the runtime based on the actor model. It adopts an SBP (\textit{split}, \textit{broadcast} and \textit{partial-value}) abstraction, 
enabling various hybrids of data parallelism and model parallelism in a much easier manner than existing frameworks. The actor model provides a succinct runtime mechanism to manage complex dependencies imposed by resource constraints, data movement and computation in distributed training.

We demonstrate the general applicability and efficiency of \SysName{} for training various large DNN models with 
extensive experiments, comparing to many representative state-of-the-art systems. 
The results show that, with a much simpler and more generic implementation, \SysName{} achieves performance comparable to or slightly better than that of the major customized libraries which are built on top of the state-of-the-art frameworks.

\vspace{-0.3cm}
\section{Background and Motivation}
\vspace{-0.2cm}

\begin{figure}
    \centering
    \includegraphics[width=\linewidth]{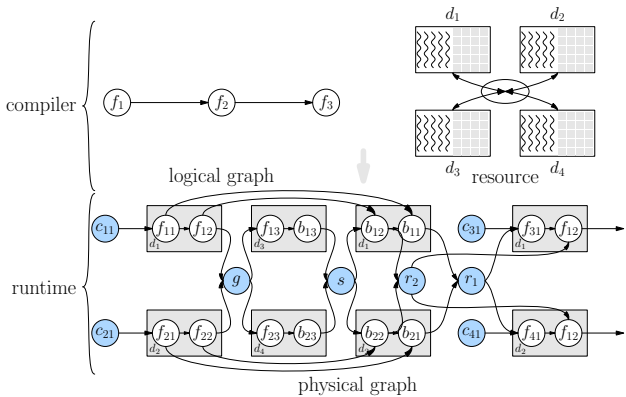}
    \vspace{-0.6cm}
    \caption{
    A typical DL framework which translates the \textit{logical} graph of a three-layer NN to a \textit{physical} graph (or execution plan) on 4 inter-connected devices. 
    }
    \label{figure:top_level_arch}
    \vspace{-0.6cm}
\end{figure}

A DNN is typically expressed as a \textit{logical} computation graph of operators (abbreviated as op) in DL frameworks, which is manually programmed or automatically converted by a \textit{compiler} into a \textit{physical} graph composed of optimized kernels for execution 
at runtime~\cite{abadi-2016-tensorflow}. 
Distributed training involves mandatory communication ops for data (gradient, parameters, or activations) exchange among devices~\cite{li14ps, priya-2017-large, jiamin-2016-corr}. 
The inter-device bandwidth is still one or two orders of magnitude lower than that of data access within a device~\cite{jiang-osdi-2020, narayanan-2019-sosp}. 
Therefore, a distributed DL framework 
should treat data movement as a first-class citizen as computation.


\vspace{-0.3cm}
\subsection{Distributing the Workload in Spatial Domain}
\vspace{-0.2cm}




\textit{Spatial Scheduling} specifies how to spread the ops across multiple devices. 
Figure \ref{figure:top_level_arch} illustrates a training job with three computation ops $f_{1},f_{2},f_{3}$ scheduled onto four inter-connected devices $d_{1},d_{2},d_{3},d_{4}$. 
$f_{1}$ and $f_{2}$ are 
executed on $d_{1}$ and $d_{2}$ with data parallelism, and $f_{3}$ runs on $d_{3}$ and $d_{4}$ with model parallelism. 
An \textit{all-gather} 
communication op $g$ is inserted between $\{f_{12},f_{22}\}$ and $\{f_{13}, f_{23}\}$ in the forward pass, while a \textit{reduce-scatter} 
communication op $s$ 
is required between $\{b_{13},b_{23}\}$ and $\{b_{12}, b_{22}\}$ in the backward pass. Two \textit{all-reduce} collective communication ops $r_{1}$ and $r_{2}$ are used to
synchronize model parameters of $f_{1}$ and $f_{2}$. 
Manually arranging the communication ops in such hybrid parallelism case by case is labor-intensive, 
incurring significant obstacles in applying complex parallelism to new DL models. 

\vspace{-0.3cm}
\subsection{Distributing the Workload in Temporal Domain}
\vspace{-0.2cm}

\textit{Temporal Scheduling} of dataflow in a DL job refers to 
scheduling execution of ops in a particular order to maximize hardware utilization and system throughput. 
The best opportunity for performance improvement usually comes from overlapping communication and computation whenever possible. Execution dependencies 
are enforced within and across different instances (each mini-batch corresponds to an instance) on a physical graph when using synchronous stochastic gradient descent training~\cite{jiamin-2016-corr}. In Figure \ref{figure:top_level_arch}, for example, forward ops $f_{31}$ and $f_{41}$ cannot be scheduled ahead of the \textit{all-reduce} op $r_{1}$. On the other hand, data loading and pre-processing ops $c_{31}$ and $c_{41}$ can be performed simultaneously while the devices are processing the previous batch of data; back-propagation $\{b_{11}, b_{21}\}$ and the \textit{all-reduce} op $r_{2}$ can be executed in parallel, without hampering the correctness. 

\vspace{-0.3cm}
\subsection{Managing the Complex Dependencies}
\vspace{-0.2cm}
In mainstream DL frameworks, 
both \textit{data} and \textit{control} dependencies are represented with edges in the execution graph~\cite{abadi-2016-tensorflow, paszke-2019-pytorch, chen-2015-mxnet}. Upon the completion of each op, the scheduler updates dependencies of the remaining ops and identifies ops that are ready to run (whose dependencies have all been resolved). Distributed DL 
often experiences increased complexity of execution dependencies and resource constraints~\cite{rajbhandari-zero-2020, huang-2019-nips}.


\begin{figure}
    \centering
    \includegraphics[width=0.9\linewidth]{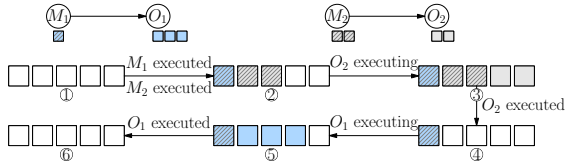}
    \caption{An example where deadlock may result with the scheduler in existing frameworks.}
    \label{figure:graph_executor}
    \vspace{-0.6cm}
\end{figure}
\noindent\textbf{Dependencies caused by resource sharing.} The scheduler has to decide an appropriate execution order to avoid out-of-memory (OOM) errors or deadlocks when multiple ops share the same resource. Consider a simple example in Figure~\ref{figure:graph_executor}. 
$M_{1}$ and $M_{2}$ are two data movement ops serving two computing ops $O_{1}$ and $O_{2}$ on the same device, respectively. $O_{1}$ and $O_2$ do not depend on each other and $O_1$ requires more device memory to execute than $O_2$. $M_{1}$ and $M_{2}$ also need some device memory to store the output data. 
After $M_{1}$ and $M_{2}$ have occupied their memory, 
the free memory capacity can only satisfy 
$O_2$ but not 
$O_1$, while both $O_1$ and $O_2$ are in the ready set of the scheduler (as in TensorFlow's) at the same time. If $O_1$ is scheduled first, the memory is insufficient; the system may either report an OOM error or block the scheduling thread, and the latter may cause a deadlock. To avoid this risk, it is better for the framework to 
specify an appropriate execution order in advance (e.g., adding \textit{control} dependencies between ops in TensorFlow). If the system leverages pipelining to overlap data movement and computation, 
the issue becomes even more severe, as $M_{1}$ can execute simultaneously while $O_{1}$ is processing the previous piece of data in the above example.
Resource planning at compile-time and flow control at runtime are necessary for 
execution stability. 

\begin{figure}
    \centering
    \includegraphics[width=0.8\linewidth]{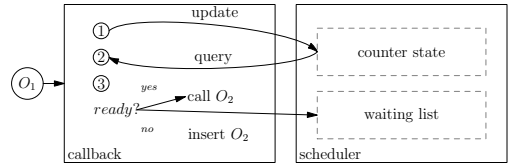}
    \caption{Interaction between callback function and the scheduler.}
    \label{figure:data_movement}
    \vspace{-0.6cm}
\end{figure}

\noindent\textbf{Dependencies caused by data movement.} The existing DL frameworks do not treat data movement (e.g., between host and device memories) as a normal op in the graph. 
As a result, the dependencies between data movement and computation are not represented with edges in the computation graph. For example, TensorFlow wraps intra-node data movements in callback functions and inserts them 
where necessary. As a result, some dependencies are represented by graph edges while others are described by the invocation of callback functions. In Figure~\ref{figure:data_movement}, $O_{2}$ is wrapped in a callback function which is expected to be invoked on the completion of $O_{1}$. However, if $O_{2}$ has other dependencies such as the output of other ops or \textit{control} dependencies, the completion of $O_{1}$ does not suffice to invoke $O_{2}$. 
To correctly schedule $O_{2}$, the callback function should tell the scheduler the completion of $O_{1}$: 
if the scheduler returns that all the other dependencies have been resolved, $O_{2}$ can be scheduled immediately; otherwise, $O_{2}$ is inserted into a waiting list and will be scheduled in the future when other dependencies are resolved. 


In the above example, the framework has to expose the internal scheduler to users so that the inserted callback functions can correctly interact with the scheduler. However, substantial engineering efforts are required to modify the existing DL frameworks to achieve this, 
as none of the existing DL frameworks expose the underlying scheduler to users yet. Ideally, the framework should represent all the dependencies among all the ops (including data movement) explicitly in the graph. Once this is achieved, 
the graph executor at runtime can also be  greatly simplified.

\vspace{-0.3cm}
\subsection{Summary}
\vspace{-0.2cm}
We design \SysName{}, with a compiler that can automatically generate a physical graph for data parallelism, model parallelism and pipeline parallelism. The compiler supports a full analysis of all types of dependencies (e.g., resource, data movement and computation) at compile-time. Furthermore, we design a succinct runtime for \SysName{} based on actor model, which instantiates all types of dependencies with a unified approach of message passing among actors.

\vspace{-0.3cm}
\section{The Compiler}\label{sec:sbp}
\vspace{-0.2cm}

\SysName{}'s compiler takes a logical computation graph and the assigned hardware configuration as inputs and generates a physical graph describing the actual execution procedure. We assume each logical op is already assigned with an attribute \textit{placement}, indicating on which nodes (i.e., physical machines) and devices the logical op will be deployed. Consequently, a global tensor (i.e., the input or the output of a logical op) is also mapped to multiple local tensors (i.e., the multiple correspondences on the devices where the logical op is placed). 

\vspace{-0.3cm}
\subsection{Specifying Parallelism of Each Tensor and Each Operator among Assigned Devices} 
\vspace{-0.2cm}

\begin{figure}[!t]
    \centering
    \includegraphics[width=0.8\linewidth]{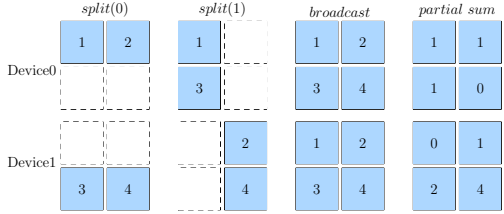}
    \caption{Example of 4 \textit{SBP} signatures to map a $2\times 2$ global tensor to two devices. Each block in the figure indicates an entry of a tensor.
    }
    \label{figure:sbp}
    \vspace{-0.3cm}
\end{figure}

We design \textit{SBP}, a mathematical abstraction specifying the mapping between a global tensor and the corresponding local tensors, including 
\textit{split} (\textit{S} in short), \textit{broadcast} (\textit{B}) and \textit{partial-value} (\textit{P}). 
The example in Figure \ref{figure:sbp} demonstrates how a global tensor with a shape of $2\times 2$ is mapped to 2 local tensors under 4 types of \textit{SBP} mappings (each referred to as an \textit{SBP} signature), namely \textit{split(0)}, \textit{split(1)}, \textit{broadcast}, and \textit{partial-sum}. \textit{split} indicates that the local tensors are obtained by splitting the global tensor along a certain axis in a balanced manner. For example, the two tensors in the first column in Figure \ref{figure:sbp} are obtained by splitting the global $2\times 2$ tensor by row axis, while the two tensors in the second column are resulted in by splitting the global tensor by column axis. As shown by the third column of Figure \ref{figure:sbp}, \textit{broadcast} means that each local tensor is an exact copy of the global tensor. As demonstrated by the last column of Figure \ref{figure:sbp}, \textit{partial-value} indicates that the local tensors have the same shape as the global tensor, and the global tensor can be obtained by performing an element-wise reduction operation (e.g., \textit{sum}, \textit{max}, etc.) over all the local tensors.  

When \textit{SBP} signatures of the input tensors of an op are given, \textit{SBP} signature of its output tensor can also be determined. Take $MatMul$ as an example. Given a data tensor $X$ and a weight tensor $W$, \textit{SBP} signature of their product $Y=XW$ can be inferred from those 
of $X$ and $W$, as given in Table \ref{tab:matmul-sbp}. For most operators, the rule for inferring the \textit{SBP} of output tensor from the \textit{SBP} of input tensors is straightforward. Take the first case in Table \ref{tab:matmul-sbp} as an example, if $X$ is split by row (i.e., $S(0)$) and $W$ is \textit{broadcast}, the result $Y$ will also be split by row (i.e., $S(0)$). Currently, we provide the \textit{SBP} deduction rule for all the operators case by case and expect to automate the process in the future. 
With \textit{SBP} signatures of an op's inputs and outputs, the parallelism strategy of the op is fully specified. For example, 
$S(0),B$ for $X,W$ in the first row of Table \ref{tab:matmul-sbp} correspond to data parallelism, and $B,S(1)$ for $X,W$ in the second row indicates model parallelism.

\begin{table}[!t]
    \caption{Valid \textit{SBP} signatures for \textit{MatMul}}
    \label{tab:matmul-sbp}
    \centering
    {\small
    \begin{tabular}{|c|c|c|}
    \hline
         $X$ & $W$ & $Y=XW$\\
    \hline
         $S(0)$ & $B$ & $S(0)$\\
    \hline
         $B$ & $S(1)$ & $S(1)$\\
    \hline
         $S(1)$ & $S(0)$ & $P(\mathit{sum})$\\
    \hline
         $P(\mathit{sum})$ & $B$ & $P(\mathit{sum})$\\
    \hline
         $B$ & $P(\mathit{sum})$ & $P(\mathit{sum})$\\
    \hline
         $B$ & $B$ & $B$\\
    \hline
    \end{tabular}}
    \vspace{-0.3cm}
\end{table}

\vspace{-0.3cm}
\subsection{Modeling Data Routing} \label{sec:boxing}
\vspace{-0.2cm}

\begin{figure}[!t]
    \centering
    \includegraphics[width=0.95\linewidth]{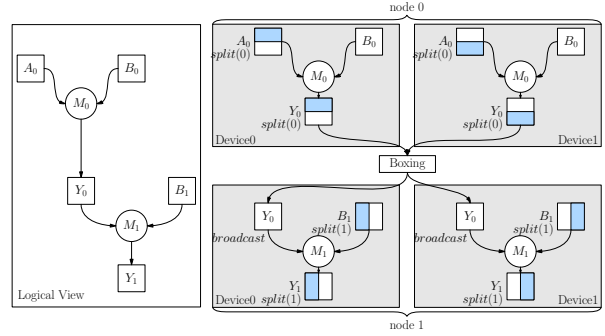}
    \vspace{-0.2cm}
    \caption{Example showing data movement with a boxing op inserted, when translating a logical graph into a physical graph. 
    }
    \label{figure:sbp_translation}
    \vspace{-0.3cm}
\end{figure}

Producer and consumer of the same global tensor may prefer different \textit{SBP} signatures for the tensor. As illustrated in Figure \ref{figure:sbp_translation}, two $MatMul$ ops are connected by a global tensor $Y_0$. $S(0)$ is $Y_{0}$'s inferred \textit{SBP} signature by $MatMul_{0}$; however, $MatMul_{1}$ expects its \textit{SBP} signature to be $B$. In this case, a data-routing op for re-arranging or transforming the local tensors of $Y_{0}$ is required between $MatMul_{0}$ and $MatMul_{1}$. In distributed DL, the data-routing op for automatically transforming the intermediate local tensors is usually one of  the common collective communication primitives such as \textit{all2all}, \textit{broadcast}, \textit{reduce-scatter}, \textit{all-reduce}, \textit{all-gather}, etc. We unify all such ops as a type of \textit{boxing} ops. 
In the example of Figure~\ref{figure:sbp_translation}, the \textit{boxing} op performs an \textit{all-gather} operation internally.

\begin{table}[!t]
    \caption{Data size transferred between successive \textit{SBP} signatures. $p_1$ ($p_2$) is the number of devices where input (output) tensors are placed. $|T|$ is 
    the size of the global tensor $T$. }
    \label{tab:transfer-cost}
    \centering
    {\small
    \begin{tabular}{|c|c|c|}
    \hline
         $\mathit{SBP}_1\!\rightarrow\! \mathit{SBP}_2$ & Cost (same) & Cost (disjoint) \\
    \hline
         $S(i)\rightarrow S(i)$& $0$ & $|T|$\\
    \hline
          \makecell[c]{$S(i)\rightarrow S(j)$ \\$(i\neq j)$}&  \makecell[c]{$\frac{p_1-1}{p_1}|T|$ \\\textit{all2all}} & $|T|$\\
    \hline
        $S\rightarrow B$&  \makecell[c]{$(p_1-1)\cdot |T|$\\ \textit{all-gather} }& $p_2\cdot |T|$ \\
    \hline
        $S\rightarrow P$ & $0$ & $|{T}|$ \\
    \hline
        $B\rightarrow S$ & $0$ & $|{T}|$ \\
    \hline
        $B\rightarrow B$ & $0$ & $p_2 \cdot|{T}|$ \\
    \hline
        $B\rightarrow P$ & $0$  & $|{T}|$ \\
    \hline
        $P\rightarrow S$ &  \makecell[c]{$(p_1-1)\cdot |{T}|$\\ \textit{reduce-scatter} }& $p_1\cdot|{T}|$ \\
    \hline
        $P\rightarrow B$ &  \makecell[c]{$2(p_1-1)\!\cdot\! |{T}|$\\ \textit{all-reduce}} & $(p_1\! +\! p_2 -1 )\cdot |{T}|$ \\
    \hline
        $P\rightarrow P$ & $0$ & $p_1\cdot |{T}|$\\
    \hline
    \end{tabular}}
\end{table}

The inserted \textit{boxing} op may or may not incur communication cost. Table~\ref{tab:transfer-cost} lists the data size transferred between successive \textit{SBP} signatures, when the input tensors and the output tensors of the \textit{boxing} op are on the same set or disjoint sets of devices, respectively.   
Tensor 
transformation across disjoint sets of devices always incurs communication costs, while 
tensor transformation within the same set of devices may not necessarily lead to data movement (e.g., $B\!\rightarrow\! S$ in Table \ref{tab:transfer-cost}, since the output tensor can be directly obtained from the input tensor located at the same device). 
This is useful for 
deciding the optimal parallelism strategy, that is, by selecting \textit{SBP} signatures incurring the lowest communication costs. 

\vspace{-0.3cm}
\subsection{Difference from GShard's Abstractions}\label{section:gshard}
\vspace{-0.2cm}

Our 
\textit{SBP} abstractions bear some similarities to those in GShard~\cite{dmitry-2020-gshard},\footnote{\textit{SBP} and GShard are independently developed being unaware of each other, which can be proved by tracking the commit logs of \SysName{} in GitHub.} i.e., \textit{split} (\textit{split} in GShard) and \textit{broadcast} (\textit{replicate} in GShard). GShard further adds a \textit{shard} annotation to generalize \textit{split} to multi-dimensional \textit{split}. In \SysName{}, we use multi-dimensional \textit{split} that unifies the \textit{split} and \textit{shard} in GShard. Besides \textit{split}, we also generalize all other \textit{SBP} signatures to multi-dimension. For example, a matrix can has an \textit{SBP} signature as $(S(0),B)$, in which $S(0)$ specifies the parallelism strategy at the level of nodes while $B$ indicates the parallelism strategy among devices inside the same node. As the deduction rule shown in Figure \ref{tab:2d-matmul-sbp}, with multi-dimensional \textit{SBP}, more advanced distributed matrix multiplication such as 2D SUMMA algorithm~\cite{xu2021efficient} can be conveniently supported.







Further, we create the \textit{partial-value} signature which GShard does not consider, 
but is necessary to make the annotation system complete. For example, Table \ref{tab:matmul-sbp} lists all the valid SBP signatures for a matrix multiplication op ($Y=XW$). If $X$ uses \textit{S(1)} and $W$ uses \textit{S(0)}, the signature of $Y$ will be \textit{P(sum)}, which cannot be described by either \textit{split} (i.e., \textit{split} and \textit{shard} in GShard) or \textit{broadcast} (i.e., \textit{replicate} in GShard). GShard suggests performing \textit{reduce} to combine the partial data to obtain the final result immediately after the un-reduced data are generated.
However, sometime, maintaining the intermediate result as the \textit{partial-value} is more efficient than immediately reducing the partial results. With 
\textit{partial-value}, \SysName{} allows the system to choose the optimal timing of inserting a \textit{boxing} op (i.e., a \textit{reduce} or \textit{all-reduce} op).  
Take $Y=U \times V \times W$ as an example. Suppose \textit{SBP} signatures of $U$, $V$ and $W$ are  \textit{S(1)}, \textit{S(0)} and \textit{B}, respectively. According to 
Table \ref{tab:matmul-sbp}, \textit{SBP} signature of the result of $U\times V$ is \textit{P(sum)}. The partial result can be multiplied by $W$, since the product of $P(sum)$ and $B$ is valid and the resulting signature is $P(sum)$. Without \textit{partial-value} signature, a \textit{boxing} op, which incurs additional communication cost, must be inserted before performing the second matrix multiplication.

\begin{table}[!t]
    \caption{Two valid two-dimensional \textit{SBP} signatures for \textit{MatMul}}
    \label{tab:2d-matmul-sbp}
    \centering
    {\small
    \begin{tabular}{|c|c|c|}
    \hline
         $X$ & $W$ & $Y=XW$\\
    \hline
         $(S(0),B)$ & $(B,S(1))$ & $(S(0),S(1))$\\
    \hline
         $(S(0),S(1))$ & $(B,S(0))$ & $(S(0),P)$\\
    \hline
    \end{tabular}}
    \vspace{-0.2cm}
\end{table}

\begin{table}[!t]
\caption{Example program for implementing SBP signatures/parallelism of $MatMul_{0}$ and $MatMul_{1}$ in Figure \ref{figure:sbp_translation}.
}
\label{tab:api_demo}
{\small
\begin{tabular} {c}
\hline 
\begin{lstlisting}[language=Python]
import oneflow as flow
P0= flow.placement("cuda", {0:[0,1]})
P1= flow.placement("cuda", {1:[0,1]})
a0_sbp=flow.sbp.split(0)
b0_sbp=flow.sbp.broadcast
y0_sbp=flow.sbp.broadcast
b1_sbp=flow.sbp.split(1)

A0=flow.randn(4,5,placement=P0,sbp=a0_sbp)
B0=flow.randn(5,8,placement=P0,sbp=b0_sbp)
Y0=flow.matmul(A0,B0) 

Y0 = Y0.to_global(placement=P1,sbp=y0_sbp)
B1=flow.randn(8,6,placement=P1,sbp=b1_sbp)
Y2=flow.matmul(Y0,B1)
\end{lstlisting}\\
\hline 
\end{tabular}}
\vspace{-0.3cm}
\end{table}

\vspace{-0.3cm}
\subsection{The Programming Interface}\label{sec:api}
\vspace{-0.2cm}










The design objective of the programming interface is to keep the operator APIs and the model description the same between a single device version and a distributed one. For different distributed strategies, users only need to specify the placement and \textit{SBP} signatures of some tensors. 
Consider the example in Figure \ref{figure:sbp_translation} where $MatMul_{0}$ and $MatMul_{1}$ use data and model parallelism, respectively. The code snippet in Table \ref{tab:api_demo} illustrates how \SysName{} achieves the respective parallelism. 
Two different placements are created in line 2 and line 3, 
where 
\textit{cuda} indicates 
NVIDIA GPGPUs as accelerators, and $\{0\!:\![0,1]\}$ and $\{\!1:\![0,1]\}$ denote 
node and device placements (the number before the colon is the node ID and numbers in square brackets are device IDs). 
\textit{SBP} signatures are created in lines 4-7. 
Lines 9, 10 and 14 specify the placement and \textit{SBP} attribute of tensor $A_{0}$, $B_{0}$ and $B_{1}$, respectively. 
In line 11, \textit{SBP} signature of $Y_{0}$ is then inferred (as $split(0)$). However, the $MatMul_{1}$ at line 15 expects the \textit{SBP} signature of $Y_{0}$ to be $broadcast$. Therefore, in line 13, 
the \emph{to\_global()} method is used to add a \textit{boxing} op between $MatMul_{0}$ and $MatMul_{1}$ as described in Section \ref{sec:boxing}, which explicitly
transforms the placement and \textit{SBP} signatures of tensor $Y_{0}$. 
In line 13, 
the \emph{to\_global()} method 
transforms the placement and \textit{SBP} signature of tensor $Y_{0}$ from $split(0)$ to $broadcast$. 
We note that, since the placements of input tensors of $MatMul_{0}$ and $MatMul_{1}$ are different, i.e., $P0$ and $P1$, respectively, the two ops actually work with pipeline parallelism.

With its APIs, \SysName{} does not require a user to 
program with various low-level communication primitives, but the user may need to specify appropriate placements and \textit{SBP} signatures for each tensor. 
Placement and parallelism strategy making entails separate in-depth investigation, as studied in \cite{jia-2019-sysml,dmitry-2020-gshard,wang-2019-eurosys,narayanan-2019-sosp,huang-2019-nips}. 
After \SysName{} integrates those strategies to automatically infer optimal placement and parallelism strategy, users will no longer manually specify the attributes of tensors or explicitly call \textit{to\_global} method.

\vspace{-0.4cm}
\section{The Runtime}
\vspace{-0.1cm}

We adopt the actor model~\cite{hewitt-ijcai-1973} in runtime design. We use an actor as a thin wrapper for each op and abstract the dependencies and resources dedicated to the op as the actor's state. Actors interact with each other through message passing instead of function invocation. 
An actor's state is updated whenever it receives a message from others. We show that the actor model can elegantly solve various issues complicated to existing DL frameworks.  


\vspace{-0.3cm}
\subsection{The Actor Model} 
\label{sec:actormodel}
\vspace{-0.2cm}



An actor in our runtime is associated with 
4 components:


\vspace{-0.2cm}
$\bullet$ \textit{Registers.} A \textit{register} is simply a container holding memory addresses of tensors. An actor is usually associated with two types of registers: \textit{in} register, used for tensors consumed by the actor, and \textit{out} register, for tensors produced by the actor. 

\vspace{-0.2cm}
$\bullet$ \textit{Messages.} Actor communicate with others by exchanging messages: a \textit{req} message from a producer (i.e., the actor generating an output) to a consumer (i.e., the actor utilizing the output), notifying the consumer a register containing newly generated tensor can be read, and an \textit{ack} message from a consumer to a producer indicating that the particular register is no longer required by the consumer. 

\vspace{-0.2cm}
$\bullet$ \textit{Actions.} An {\em action} corresponds to the execution of an op that an \textit{actor} is bound to (e.g., launching a GPU kernel or performing data movement). 

\vspace{-0.2cm}
$\bullet$ \textit{A state machine.} Each actor keeps track of whether all the dependencies 
are resolved. 
    



We next discuss 
the mechanism inside each actor's state machine 
and the message passing protocol.

\vspace{-0.3cm}
\subsection{Explicit Representation of Resource Dependency}\label{section:resource_dependency}
\vspace{-0.2cm}



\noindent\textbf{Counters for both \textit{in} and \textit{out} registers.} Each actor allocates a pre-determined number of \textit{out} registers in the beginning, amounting to a fixed memory quota for each actor. If an actor has used up its quota, the next \textit{action} will not be scheduled even all its input tensors have been ready, 
until some memory previously allocated to the actor can be recycled. To achieve such goal, we associate a counter with each register. 
The zero initialized \textit{in counter} records the number of the tensors held by an \textit{in} register which is ready to be consumed, 
while the non-zero initialized \textit{out counter} represents free memory quota. 
Each action results in a decrease of some \textit{out counter}. 
Only when the \textit{in counter} equals to an expected non-zero values  and the \textit{out counter} is non-zero (indicating it has free memory to use
), can the actor trigger an action. 

In existing DL frameworks, the scheduler considers an op can start once its input tensors are ready, without taking into account whether it can later successfully acquire memory for the output. After the op is scheduled and only just before executing the action, the runtime tries to allocate memory for the op on the fly, which, however, may succeed or not. With \textit{in counter} and \textit{out counter}, \SysName{} represents resource availability as an explicit dependency for the scheduler to decide whether an op is ready to execute. Consequently, the resource planning at compile-time and flow control at runtime are made possible.




\begin{figure}
    \centering
    \includegraphics[width=0.95\linewidth]{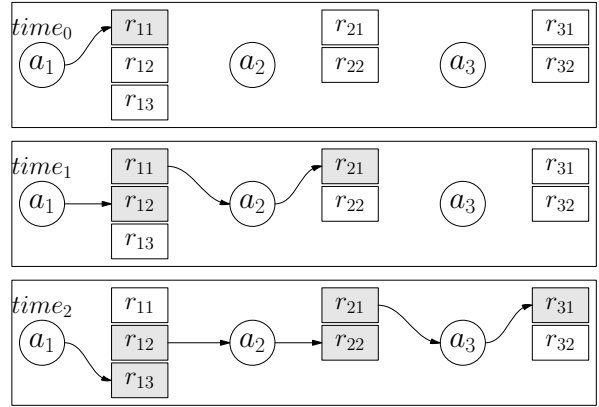}
    \caption{Pipelining example with \SysName{}'s actor-based runtime 
    A blank block indicates a register containing no useful data. 
    A filled block denotes a register 
    containing data useful to other actors. 
    }
    \label{figure:flow_control}
    \vspace{-0.5cm}
\end{figure}

\noindent\textbf{Reference counting with message passing.}
Besides the \textit{in counter} and \textit{out counter}, we introduce an additional zero-initialized \textit{reference counter} for each \textit{out} register recording the number of consumers who are referencing its content. A non-zero value of a \textit{reference counter} for an \textit{out} register indicates the register is in use and the content can not be modified. Therefore, the \textit{out counter} depends on the \textit{reference counter}. It turns out that the \textit{reference counter} can be updated according to a message passing protocol:

\vspace{-0.2cm}
$\bullet$ A producer sends a \textit{req} message to a consumer and increases the \textit{reference counter} of the \textit{out} register relating to the message by one. A change from zero to non-zero of a \textit{reference counter} results in the decrease of an \textit{out counter}. 
\vspace{-0.2cm}

$\bullet$ On receiving a \textit{req} message, the consumer knows an \textit {in} register becomes available and increases the \textit{in counter} by one.
\vspace{-0.2cm}

$\bullet$ After using data from the \textit{in} register, the consumer decreases the \textit{in counter} by one and sends an \textit{ack} message to 
the producer. 
\vspace{-0.2cm}
    
$\bullet$ On receiving an \textit{ack} message from the consumer, the producer decreases the \textit{reference counter} of the \textit{out} register relating to the \textit{ack} message, indicating the elimination of a reference on the \textit{out} register. If the \textit{reference counter} becomes zero again, the corresponding \textit{out counter} increases by one, indicating the corresponding \textit{out} register can be recycled for the future use.
In the above protocol, if an \textit{out} register is being consumed by some consumer, its \textit{reference counter} must be non-zero and it will be no longer used by the producer to put newly generated tensors. Such a mutual exclusion property safely enables a zero-copy mechanism: 
if a pair of producer and consumer reside on the same device, the consumer can just directly use the producer's output as input, without making another copy of the content as input.

\subsection{Applications: pipelining and back pressure} \label{section:pipeline} 
Allowing the initial value of an \textit{out counter} for a particular register to be larger than one facilitates the processing of different versions of data in parallel. 
Each actor runs independently, acting as a natural stage in a pipeline.  
Multiple versions of the same \textit{register} can be deemed as a generalization of the double buffering technique used in traditional DL frameworks~\cite{nvidia-dali} 
In Figure \ref{figure:flow_control},
$actor_{1}$ has 3 \textit{out} registers;
$actor_{2}$ and $actor_{3}$ have 2 \textit{out} registers respectively. 

\vspace{-0.2cm}
$\bullet$ At $time_{0}$,
$actor_{1}$ produces a register $r_{11}$,
while $actor_{2}$ and $actor_{3}$ are idle because their \textit{in counter}s are zero.

\vspace{-0.2cm}
$\bullet$ At $time_{1}$,
$actor_{2}$ triggers an action because both its \textit{in counter} and \textit{out counter} are non-zeros. 
At the same time, $actor_{1}$ and trigger an action again (on a different micro-batch) because its \textit{out counter} is still non-zero. 

\vspace{-0.2cm}
$\bullet$ At $time_{2}$, actions of all 3 actors can be triggered since all their requirements on registers are fulfilled.

Essentially, the actor-based protocol is equivalent to the credit-based flow control method in asynchronous transfer mode networks~\cite{kung-credit-1994}. It naturally enables back pressure for resource preservation. If all its \textit{out} registers
are in use, a producer stops processing due to \textit{out counter} becoming zero and no available free \textit{out} register to hold the new output tensor. Without this back pressure mechanism (as in existing frameworks), a producer may run out of memory quickly if the consumer blocks.

\vspace{-0.3cm}
\section{The Implementation}
\begin{figure}[!t]
    \centering
    \includegraphics[width=0.98\linewidth]{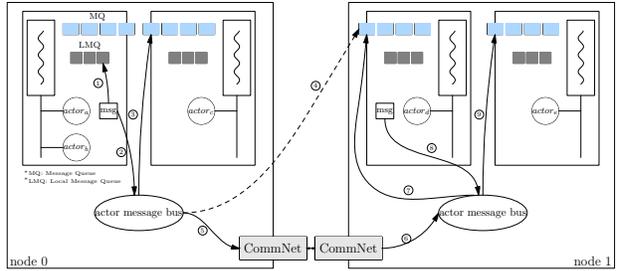}
    \vspace{-0.2cm}
    \caption{An illustration of 3 message routing cases: sending message to an actor on the same thread, sending message to an actor on another thread in the same node, and sending message to an actor on another node. The \textit{CommNet} in the figure indicates the low-level networking module in \SysName{}.
    }
    \label{fig:msg_bus}
    \vspace{-0.3cm}
\end{figure}

\begin{figure}
    \centering
    \includegraphics[width=0.98\linewidth]{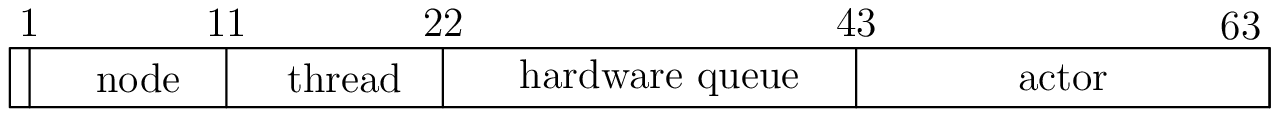}
   \vspace{-0.2cm}
    \caption{Encoding of an actor's address.}
    \label{fig:id}
    \vspace{-0.5cm}
\end{figure}

\vspace{-0.2cm}
We implement \SysName~using around 26K LoC in Python, 120K LoC in C++, and 10K LoC in CUDA. The actor runtime uses 3K LoC of C++, and the compiler module is implemented in 20K LoC of C++.\footnote{The code of \SysName{} is available at: \url{https://github.com/Oneflow-Inc/oneflow}.} In the following, we present some implementation details of actor system.
\noindent\textbf{Actor addressing and message routing.} Similar to 
CUDA stream in Nvidia GPGPUs, we also abstract other 
hardware resources (e.g., network and CPUs) as FIFO queues. We ensure no implicit dependency is brought by sharing resources. For example, two separate CUDA streams are created for copy engine and compute engine. 
To minimize 
device context switch, \SysName{} creates a dedicated OS thread for each hardware queue and the actors using the same queue (or hardware resource) are bound to the same OS thread (e.g., $actor_a$ and $actor_b$ in Figure~\ref{fig:msg_bus}). With static binding among actor, device, OS thread and node, \SysName{} assigns a unique and hierarchically organized 64-bit address (or equivalently, ID) for each 
actor as shown in Figure~\ref{fig:id}; 
 IDs of the device, OS thread and the node (where the actor resides) can be parsed from some specific fields of an actor ID.
With this ID translation mechanism, attaching the receiver actor's ID with the message suffices to route the message to its destination.





In \SysName{}, actors running on the same OS thread share a FIFO message queue. For an actor to receive a message, 
the message is first put in the message queue of the corresponding OS thread, which polls the queue repeatedly, fetches the message and routes it to the intended receiver (e.g., case \textcircled{3} in Figure~\ref{fig:msg_bus}). There is also a local message queue on each OS thread. The message sent to a receiver on the same OS thread as the sender is put into a local message queue and is directly processed by the receiver without being polled by the 
OS thread (case \textcircled{1} in Figure~\ref{fig:msg_bus}).

\noindent\textbf{Unifying the intra- and inter-node actor systems.}
We introduce an abstraction layer, the actor message bus, that provides a unified interface to route a message to its receiver no matter whether the receiver is on the same or another node. In Figure~\ref{fig:msg_bus}, the message from $actor_{a}$ to $actor_{d}$ 
travels along the logical path \{\textcircled{2},\textcircled{4}\}, while its actual path is \{\textcircled{2},\textcircled{5}, \textcircled{6}, \textcircled{7}\}. Such abstraction 
hides low-level communication across networks. 

Different from existing frameworks and libraries which insert \textit{Send} and \textit{Recv} ops at both sides of inter-node communication, 
\SysName{}'s \textit{compiler} only inserts a \textit{networking} actor at the consumer's side for pulling data from the producer's node to the consumer's node, once inter-node communication is detected. 
In Figure~\ref{fig:msg_bus}, suppose $actor_{e}$ on node 1 requires the output of $actor_{a}$ on node 0; when generating the physical graph, 
the \textit{compiler} creates $actor_{d}$ at node 1 whose sole responsibility is to pull the output of $actor_{a}$ from node 0 to node 1, so that $actor_{e}$ can consume the data as if the producer 
was on the same node.

\vspace{-0.3cm}
\section{
Evaluation}
\label{sec:evaluation}
\vspace{-0.2cm}

We demonstrate 
\SysName{}'s generality, flexibility and efficiency by 
implementing representative parallelisms 
and comparing with state-of-the-art libraries in various cases. 
Unless stated otherwise, we conduct experiments on a cluster of 4 machines inter-connected by a 100Gbps RoCE network. Each machine is equipped with 8 Nvidia Tesla V100 16G GPUs interconnected with NVLink. 

\vspace{-0.3cm}
\subsection{Data-preprocessing Pipeline}
\vspace{-0.2cm}

\begin{figure}
     \centering
         \includegraphics[width=0.9\linewidth]{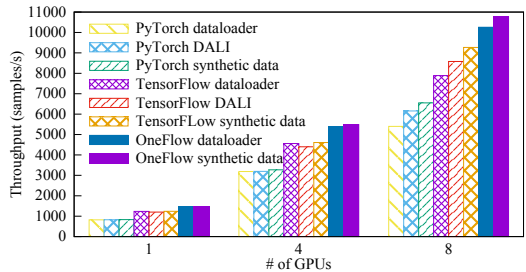}
         \vspace{-0.4cm}
         \caption{Throughput comparison with various frameworks and data loaders: training ResNet50-V1.5 with mixed precision. }
        \vspace{-0.3cm}
         \label{figure:syn_data_dali}
\end{figure}

In many scenarios such as training small DL models in mixed precision mode with high-end GPGPUs, 
feeding data to computation renders a bottleneck in DNN training \cite{kumar2020exploring}. Figure \ref{figure:syn_data_dali} compares the throughput achieved by \sys{} and mainstream frameworks with various data loaders. DALI is a 
plugin developed by Nvidia for optimizing data loading for DL frameworks~\cite{nvidia-dali}. In ``synthetic data'' cases, we use fake data generated in memory without the need for data loading from disks, representing the respective ideal cases. 
Tensorflow and PyTorch's 
data loaders are able to overlap data loading and computation but 
perform much worse than using Nvidia DALI. 
Unlike using customized plugin such as DALI, \SysName{} supports pipelining 
by just allocating two \textit{out} registers for data loading, pre-processing and copying host to device ops as described in Section \ref{section:pipeline}. Performance of \SysName{}'s data loader is 
close to that of the synthetic data case, indicating perfect piplelining between data loading actors and pre-processing actors. 
\SysName{} achieves this without additional engineering efforts like DALI.





\begin{figure*}
     \centering
     \begin{subfigure}[b]{0.245\linewidth}
         \centering
         \includegraphics[width=1\linewidth]{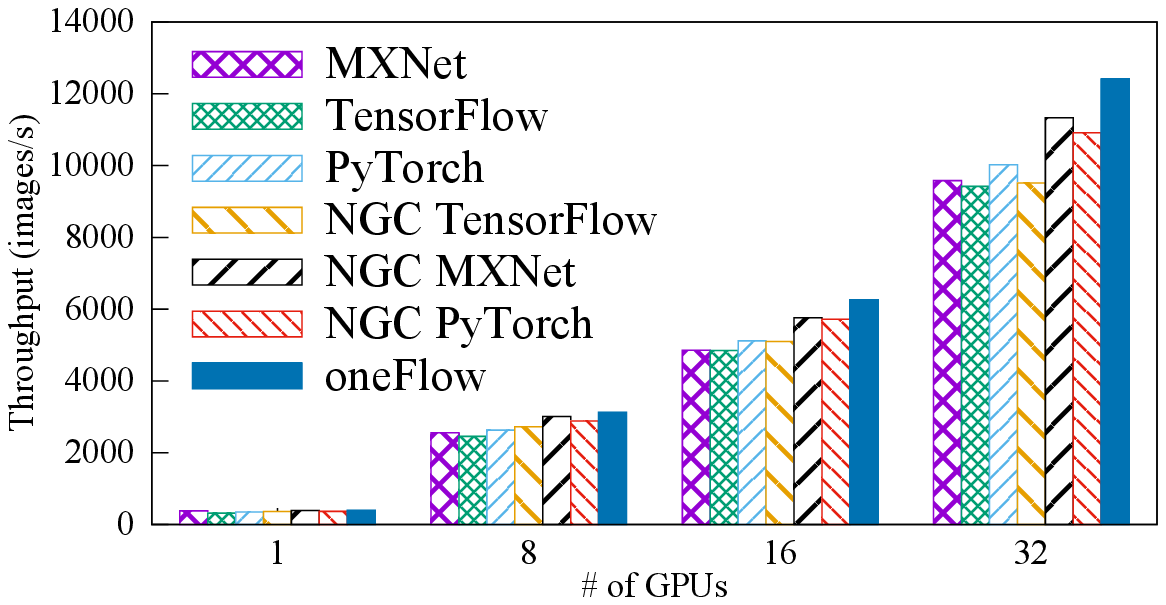}
         \vspace{-0.8cm}
         \caption{ResNet, FP32}
         \label{figure:resnet1}
     \end{subfigure}
     \begin{subfigure}[b]{0.245\linewidth}
         \centering
         \includegraphics[width=1\linewidth]{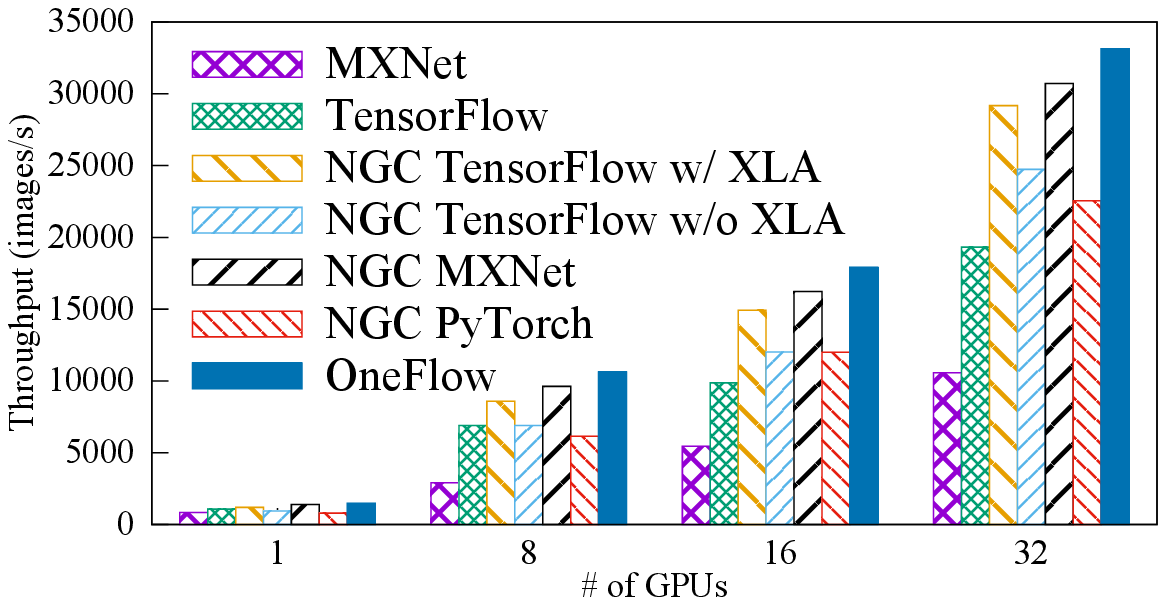}
         \vspace{-0.8cm}
         \caption{ResNet, FP16}
         \label{figure:resnet2}
     \end{subfigure}
     \begin{subfigure}[b]{0.245\linewidth}
         \centering
         \includegraphics[width=1\linewidth]{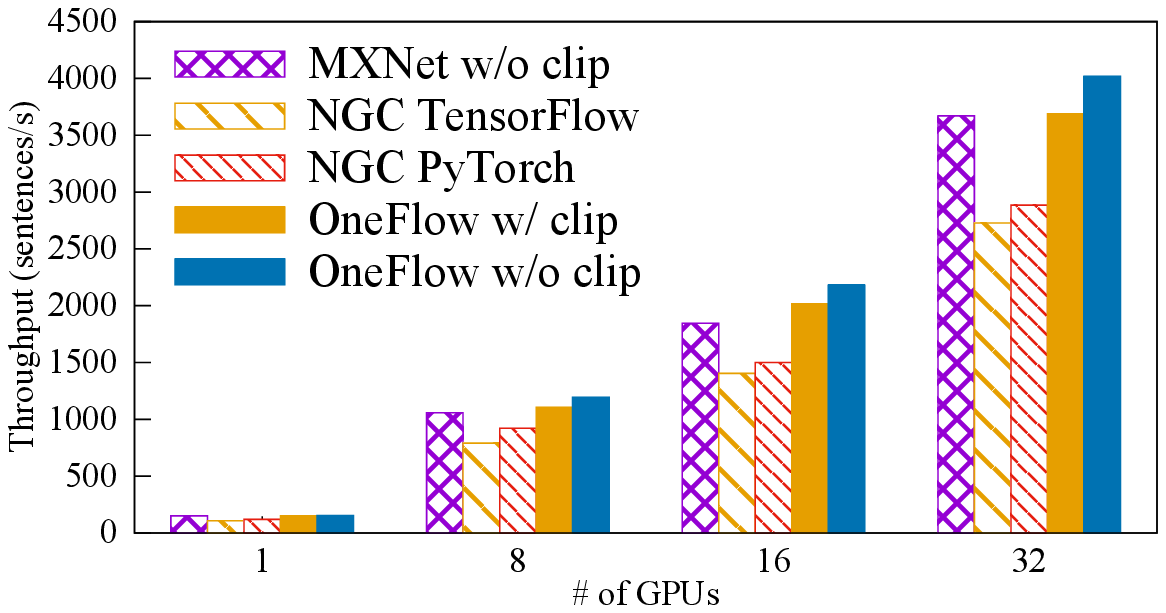}
        \vspace{-0.8cm}
         \caption{BERT-base, FP32}
         \label{figure:bert1}
     \end{subfigure}
     \begin{subfigure}[b]{0.245\linewidth}
         \centering
         \includegraphics[width=1\linewidth]{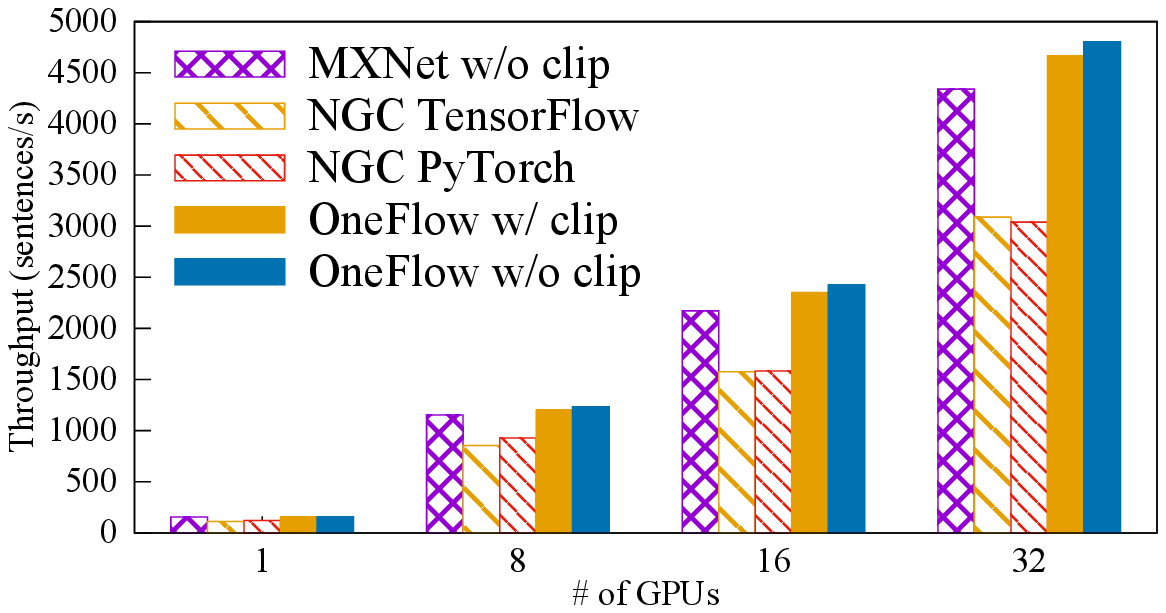}
        \vspace{-0.8cm}
        \caption{BERT-base, FP16}
         \label{figure:bert2}
     \end{subfigure}
    \vspace{-0.3cm}
     \caption{Data parallelism training of ResNet and BERT-base.}
     \label{figure:resnet-bert}
\end{figure*}






\vspace{-0.3cm}
\subsection{Data Parallelism}
\vspace{-0.2cm}

The existing DL frameworks have carried out the most extensive optimization on data-parallel training. In the experiments of Figure \ref{figure:resnet-bert}, MXNet is based on Horovod~\cite{alexander-2018-horovod}; Tensorflow and PyTorch use their native communication strategies, which lead to better performance than using Horovod. 
We observe that in the case of ResNet~\cite{he2016deep}, \SysName{} not only outperforms the official TensorFlow, PyTorch and MXNet by 23\%--31\% with FP32 and 71\%--213\% with FP16~\cite{micikevicius2018mixed}, but also outperforms the highly optimized versions of these frameworks (those prefixed by NGC, using the same script as submitted by NVIDIA to MLPerf~\cite{mlperf-training}) by 9\%--30\% with FP32 and 8\%--47\% with FP16. In terms of BERT~\cite{devlin-etal-2019-bert}, \SysName{} also achieves higher training throughput than NGC versions by 9\%--47\% with FP32 and around 55\% with FP16. For each model, we carry out a lot of performance optimization to ensure the throughput of \SysName{} on a single device comparable to or slightly better than that of other frameworks. In this way, the scalability of different frameworks can be compared based on almost the same baseline. Note that the BERT implementation in MXNet does not perform gradient clipping, which hence involves fewer computation. To perform a fair comparison between MXNet and \SysName{}, we implement two versions of BERT on \SysName{}, with and without gradient clipping, respectively. 




\begin{figure}[!t]
     \centering
     \begin{subfigure}[b]{1\linewidth}
         \centering
         \includegraphics[width=1\linewidth]{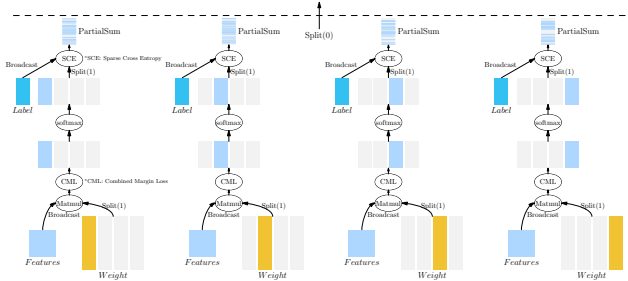}
         \caption{MatMul, softmax and sparse cross entropy operators if the \textit{SBP} signature of weight matrix is $S(1)$ .}
         \label{figure:insight_face_fc}
     \end{subfigure}\\
     \begin{subfigure}[b]{1\linewidth}
         \centering
         \includegraphics[width=1\linewidth]{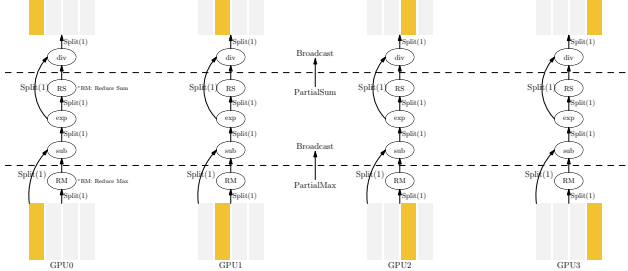}
         \caption{The details of softmax op in the physical graph generated by compiler.}
         \label{figure:insight_face_softmax}
     \end{subfigure}
         \vspace{-0.5cm}
     \caption{Implementing model parallelism in InsightFace on four GPUs.}
     \label{figure:insight}
\end{figure}

\begin{figure}[!t]
     \centering
     \begin{subfigure}[b]{0.48\linewidth}
         \centering
         \includegraphics[width=1\linewidth]{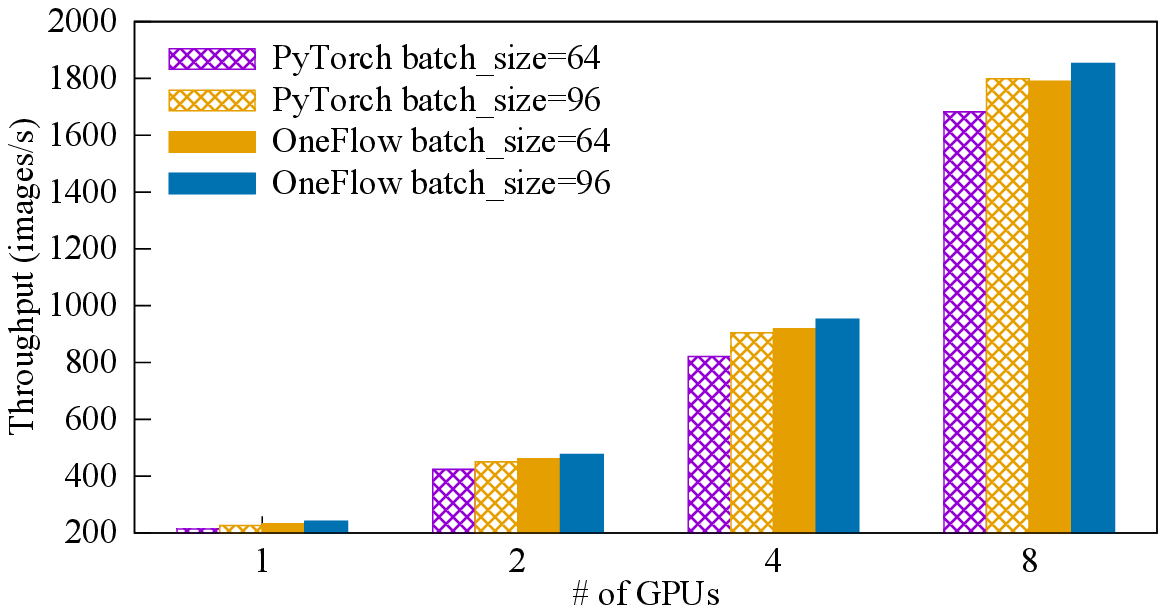}
         \caption{ResNet}
         \label{figure:insight_res}
     \end{subfigure}
     \begin{subfigure}[b]{0.48\linewidth}
         \centering
         \includegraphics[width=1\linewidth]{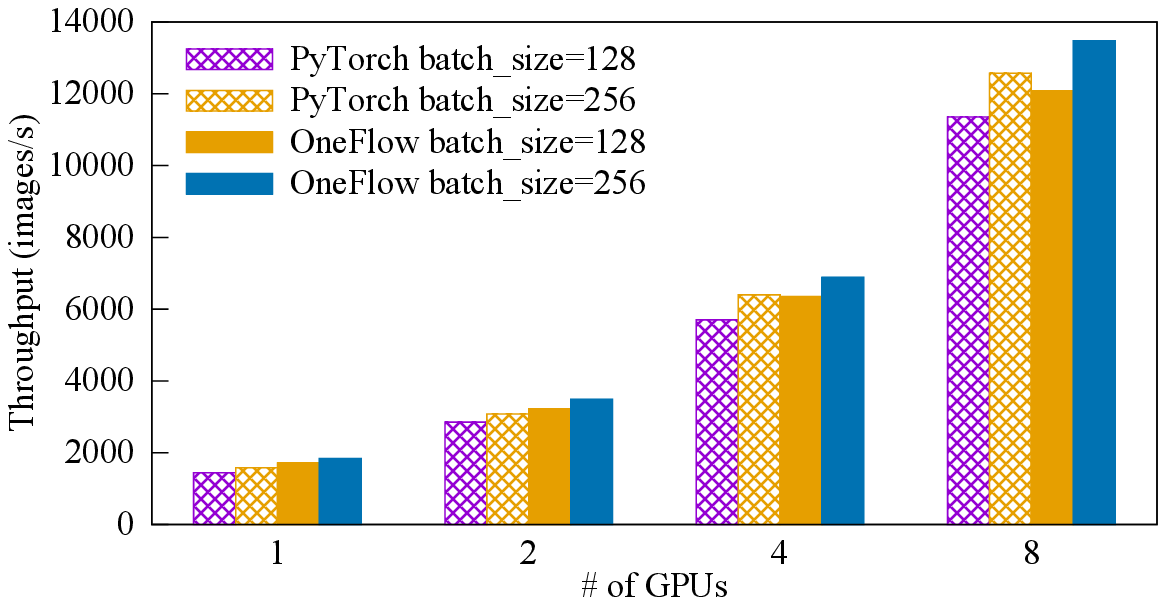}
         \caption{MobileFaceNet}
         \label{figure:insight_mobile}
     \end{subfigure}
     \vspace{-0.3cm}
     \caption{Model-parallel training
     : \SysName{} vs.~InsightFace.} 
          \vspace{-0.3cm}
     \label{figure:insight_evaluation}
\end{figure}

\vspace{-0.3cm}
\subsection{Model Parallelism}
\vspace{-0.2cm}
We compare \SysName{} with two customized DL libraries supporting model parallelism training, 
as official versions of TensorFlow and PyTorch do not support model parallelism.

\vspace{-0.3cm}
\subsubsection{InsightFace}
\vspace{-0.2cm}
InsightFace \cite{insightface} is 
widely used to train huge face recognition models, 
where model parallelism is necessary. It supports model parallelism based on PyTorch with a complicated customization.
In contrast, \SysName{} only needs to configure appropriate \textit{SBP} signatures for $MatMul$ and softmax ops that require model parallelism. Figure~\ref{figure:insight_face_fc} illustrates the transformation of local tensors on four GPUs after setting \textit{SBP} signature of the weight matrix as $S(1)$. Figure~\ref{figure:insight_face_softmax} demonstrates the details of a softmax op in the physical graph generated by the compiler. Note that, there are two \textit{reduce} calculations within the softmax op. To minimize the communication cost incurred by global reduction, \SysName{} first carries out local reduction within a device while performing the \textit{max} and \textit{sum} ops. In Figure \ref{figure:insight_evaluation}, we 
observe that \SysName{}'s throughput slightly outperforms InsightFace's when training face recognition models with ResNet and  MobileFaceNet as backbone networks respectively~\cite{chen2018mobilefacenets}. The physical execution plans used by both frameworks are essentially the same. However, the plan in InsightFace 
is generated with manual programming, while the plan in \SysName{} is automatically produced by the compiler. \SysName{} significantly eases the programming burden of model parallelism. 




\begin{figure}[!t]
     \centering
     \begin{subfigure}[b]{0.48\linewidth}
         \centering
         \includegraphics[width=1\linewidth]{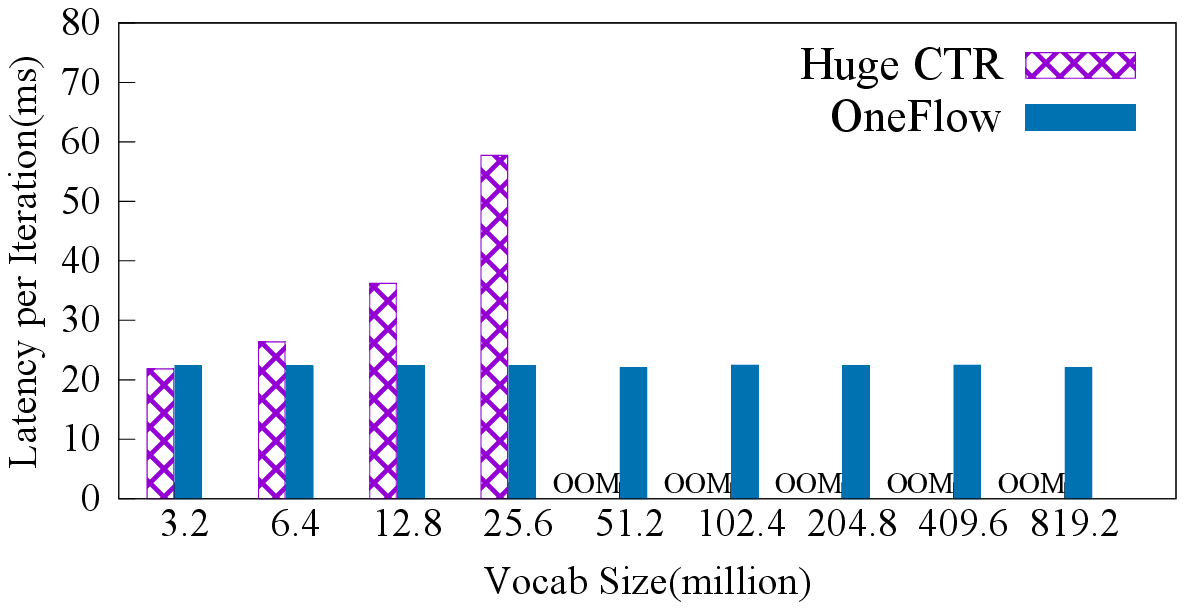}
         \caption{Training time per iteration}
     \end{subfigure}
     \begin{subfigure}[b]{0.48\linewidth}
         \centering
         \includegraphics[width=1\linewidth]{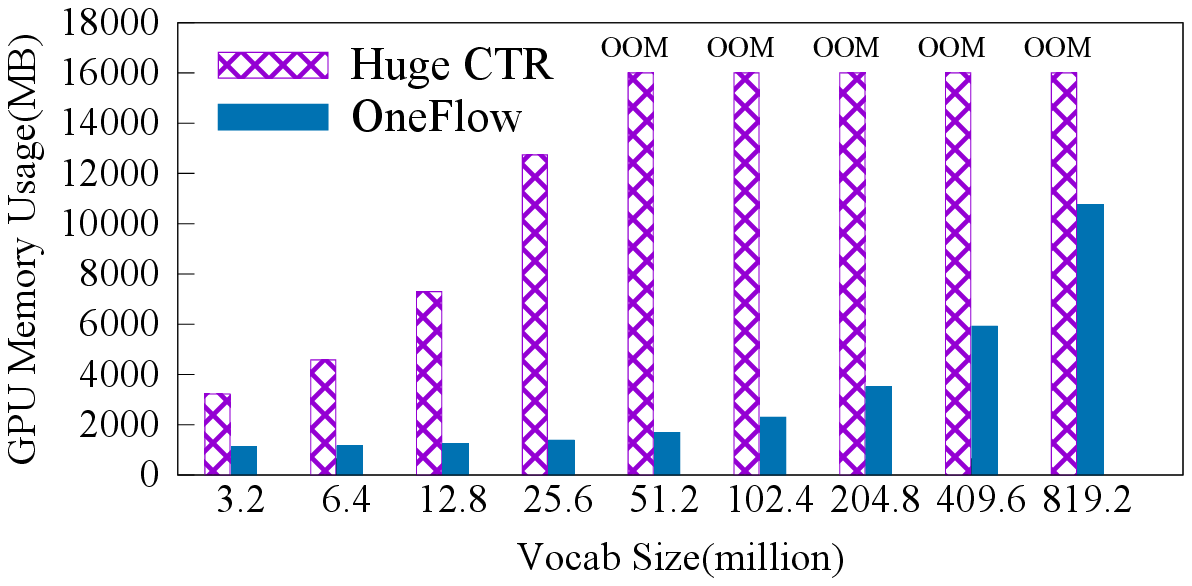}
         \caption{Memory footprint}
     \end{subfigure}
     \caption{Model parallelism training: \SysName{} vs.~HugeCTR. 
     }
          \vspace{-0.3cm}
     \label{figure:hugectr_of_latency_memory}
\end{figure}

\vspace{-0.3cm}
\subsubsection{HugeCTR}
\vspace{-0.2cm}

Wide \& Deep Leaning \cite{cheng2016wide} is widely used in recommender systems, e.g., for click-through rates estimation. In production, to support click-through rates estimation for billions of IDs, the embedding matrices become too large for a single GPU's memory to hold. Model parallelism on the embedding table is needed. As shown in Figure \ref{figure:hugectr_of_latency_memory}, Wide \& Deep learning built on \sys{} outperforms HugeCTR in NVIDIA Merlin \cite{Oldridge2020MerlinAG}, a dedicated framework for training click-through rates estimation model with model parallelism. \SysName{}achieves lower latency (i.e., per-iteration training time)  
and less memory footprint compared to HugeCTR. HugeCTR runs out of memory when the vocabulary size exceeds 51.2 million. Different from HugeCTR that involves substantial engineering efforts, with \SysName{}, users only need to set appropriate \textit{SBP} signatures for the embedding table (i.e., $S(0)$ for splitting the vocabulary IDs and $S(1)$ for splitting the hidden dimension)
, and our compiler will automatically insert collective communication ops where necessary for model parallelism. 

\vspace{-0.3cm}
\subsection{Parallelizing the Optimizer}
\vspace{-0.2cm}

\begin{figure}[!t]
     \centering
     \begin{subfigure}[b]{1\linewidth}
         \centering
         \includegraphics[width=1\linewidth]{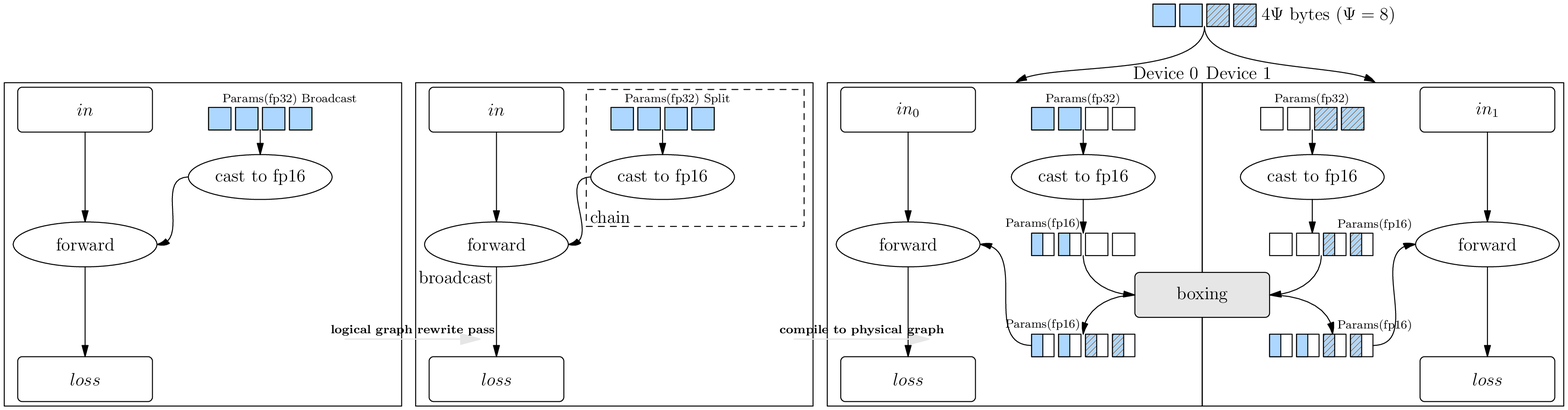}
         \caption{Zero optimizer foward pass.}
         \label{figure:zero_optimizer}
     \end{subfigure}\\
     \begin{subfigure}[b]{1\linewidth}
         \centering
         \includegraphics[width=1\linewidth]{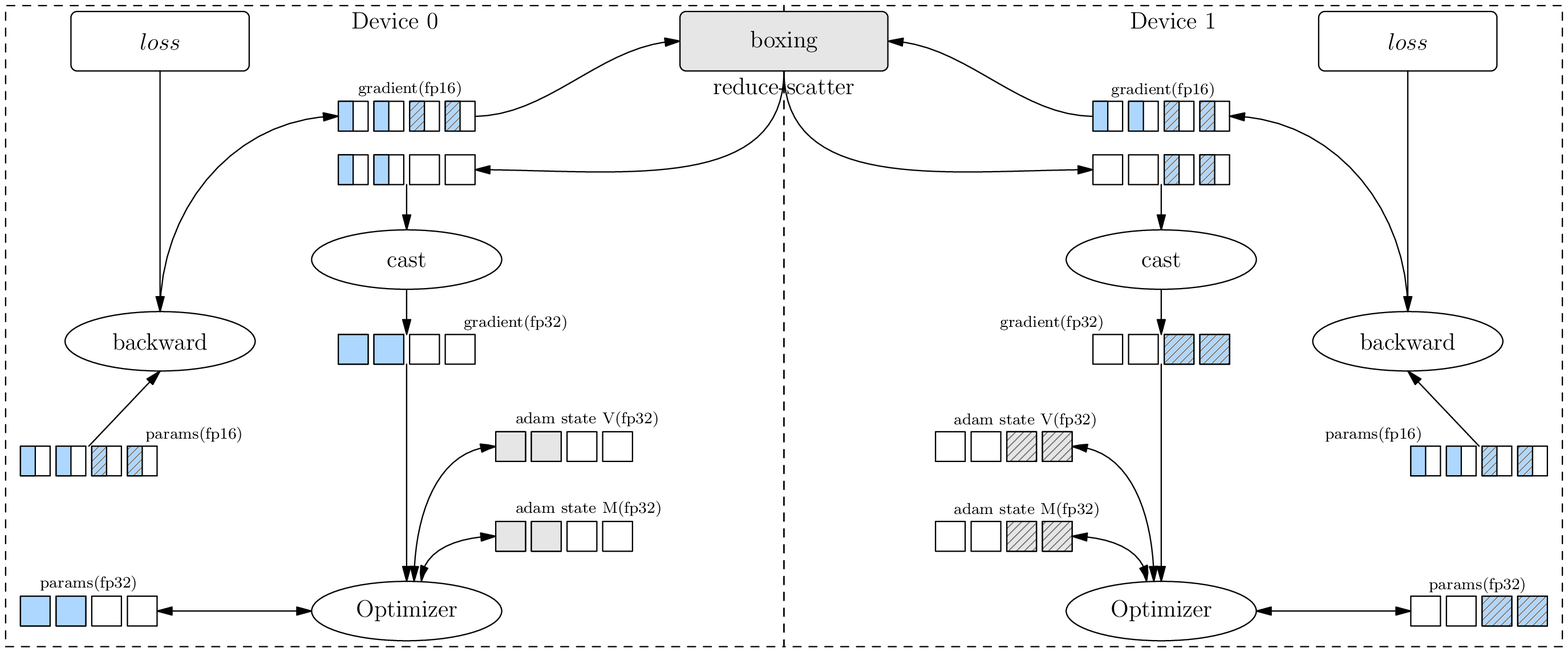}
         \caption{Zero optimizer backward pass.}
         \label{figure:zero_optimizer_backward}
     \end{subfigure}
     \vspace{-0.3cm}
     \caption{Parallelizing the optimizer in \sys{}
     .}
     \vspace{-0.3cm}
     \label{figure:zero}
\end{figure}

Memory redundancy of model states (such as gradients, parameters, momentum and variances in Adam~\cite{kingma2014adam}) in data parallelism can be significantly reduced by sharding them across devices. \textit{ZeRO}-DP \cite{rajbhandari-zero-2020} leverages it 
to support distributed training of large models on devices with limited memory, with each device only holding part of the sharded model states. When the full model states are required, an \textit{all-gather} 
communication primitive can be used. \SysName{} is able to implement the same idea with less engineering efforts. Figure~\ref{figure:zero} illustrates the procedure of generating the physical graph on two devices by \SysName{}, while implementing the same techniques as in \textit{ZeRO}-DP with mixed precision enabled~\cite{micikevicius2018mixed}. First, a conversion op (such as \textit{fp}16 cast) is inserted. Second, our framework configures \textit{SBP} signatures of the input of the cast op as $S(0)$ and the output of the cast op as $B$. Our compiler 
automatically generates the physical graph for both forward pass (Figure~\ref{figure:zero_optimizer}) and backward pass (Figure~\ref{figure:zero_optimizer_backward}). Data routing ops are automatically inserted where appropriate.
\emph{ZeRO}-DP's implementation is based on PyTorch, using about 2K LoC. 
\SysName{} implements the idea with 300 LoC, which is much simpler. 

\begin{figure}[!t]
     \centering
     \begin{subfigure}[b]{0.48\linewidth}
         \centering
         \includegraphics[width=1\linewidth]{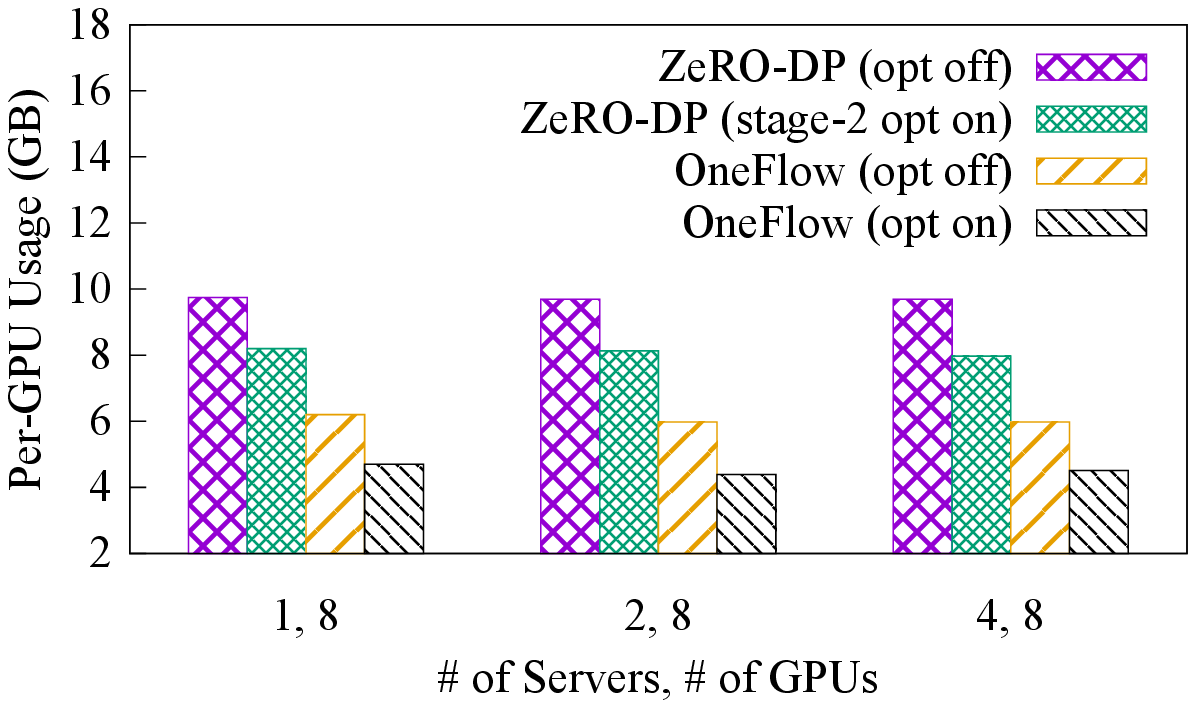}
         \caption{Memory footprint}
     \end{subfigure}
     \begin{subfigure}[b]{0.48\linewidth}
         \centering
         \includegraphics[width=1\linewidth]{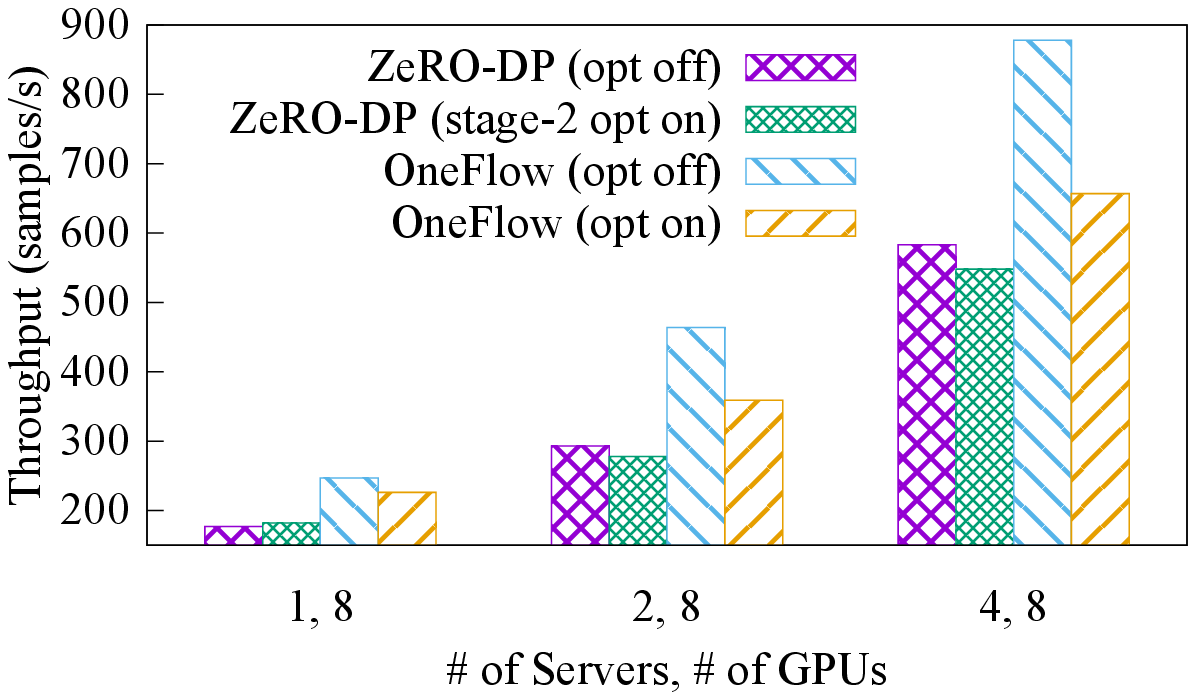}
         \caption{Throughput}
     \end{subfigure}
     \caption{
     Performance of optimizer sharding: \SysName{} vs.~
     \emph{ZeRO}-DP. 
     }
     \label{figure:deep_speed}
     \vspace{-0.5cm}
\end{figure}

Figure \ref{figure:deep_speed} compares per-device memory footprint and throughput when training GPT-2, 
with the activation checkpoint \cite{chen2016training} on (i.e., opt on) or off (i.e., opt off). We observe that \SysName{} consumes less device memory but achieves higher throughput than \emph{ZeRO}-DP, with or without the activation checkpointing optimization. 

\vspace{-0.3cm}
\subsection{Hybrid Parallelism}
\begin{figure}
     \centering
     \begin{subfigure}[b]{0.48\linewidth}
         \centering
         \includegraphics[width=1\linewidth]{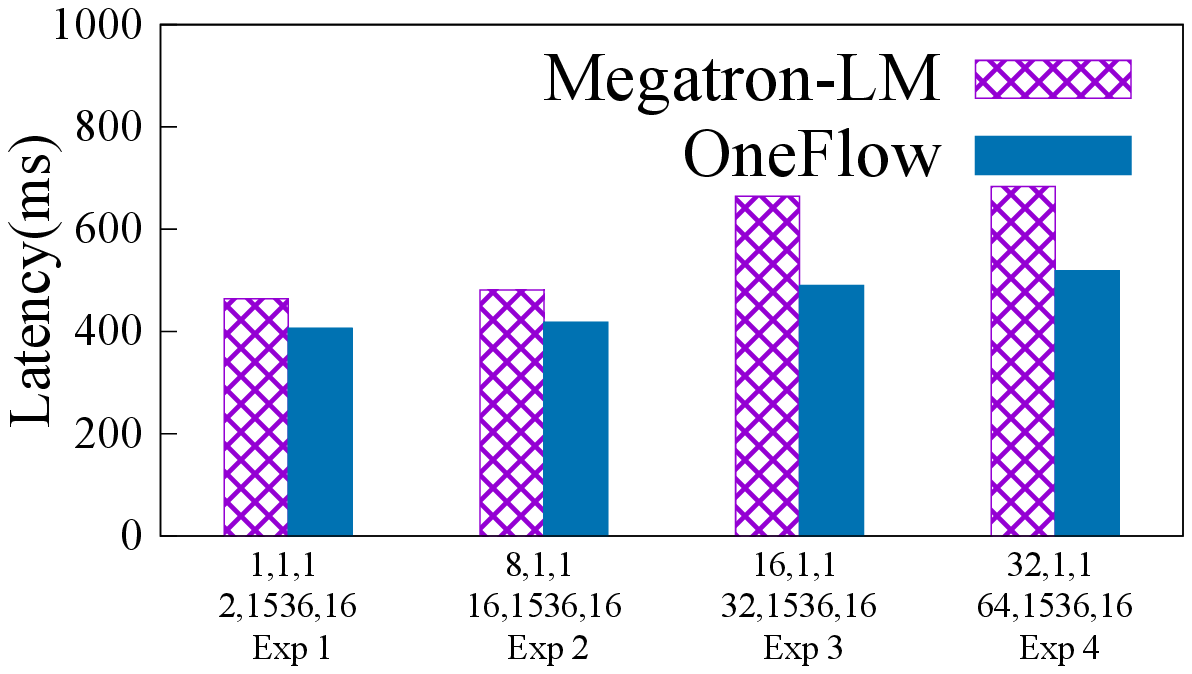}
         \caption{DP}
     \end{subfigure}
     \begin{subfigure}[b]{0.48\linewidth}
         \centering
         \includegraphics[width=1\linewidth]{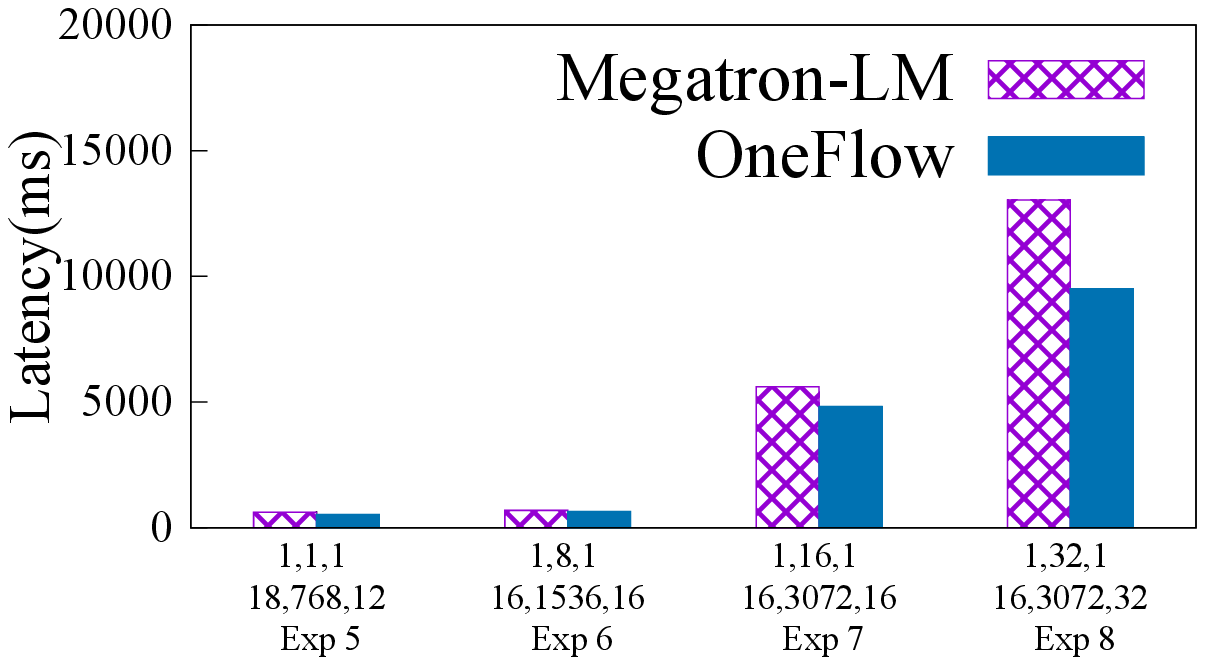}
         \caption{MP}
     \end{subfigure}
     \begin{subfigure}[b]{0.48\linewidth}
         \centering
         \includegraphics[width=1\linewidth]{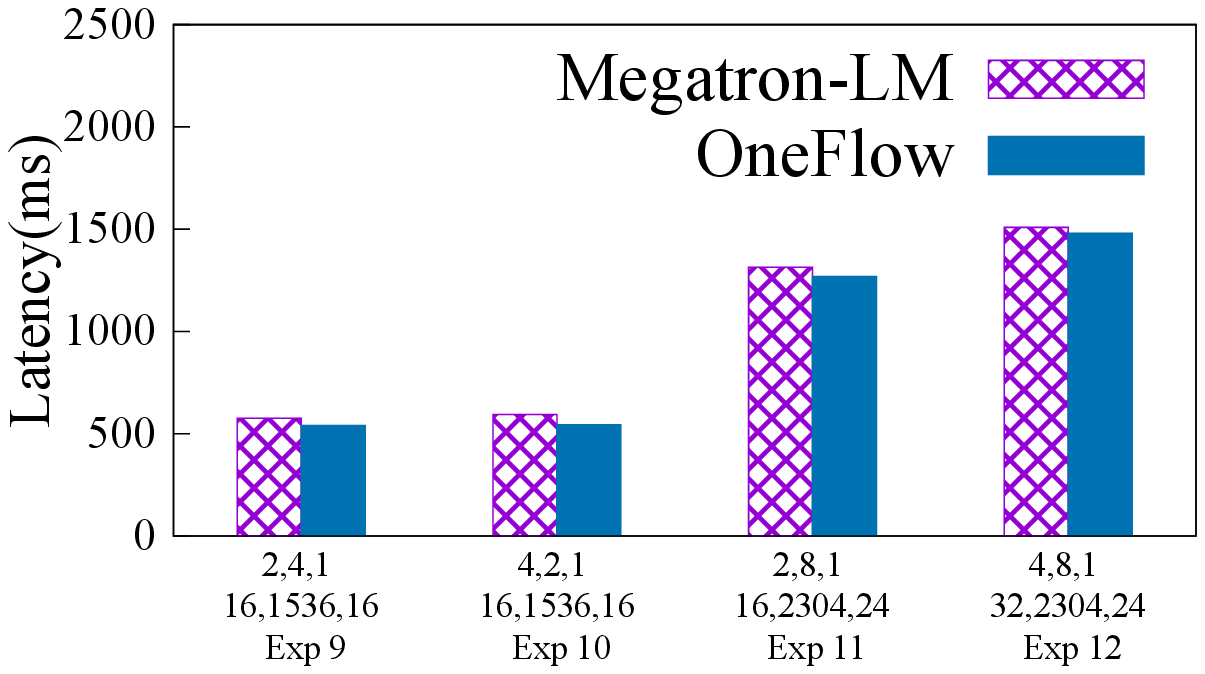}
         \caption{DP combined with MP}
     \end{subfigure}
     \begin{subfigure}[b]{0.48\linewidth}
         \centering
         \includegraphics[width=1\linewidth]{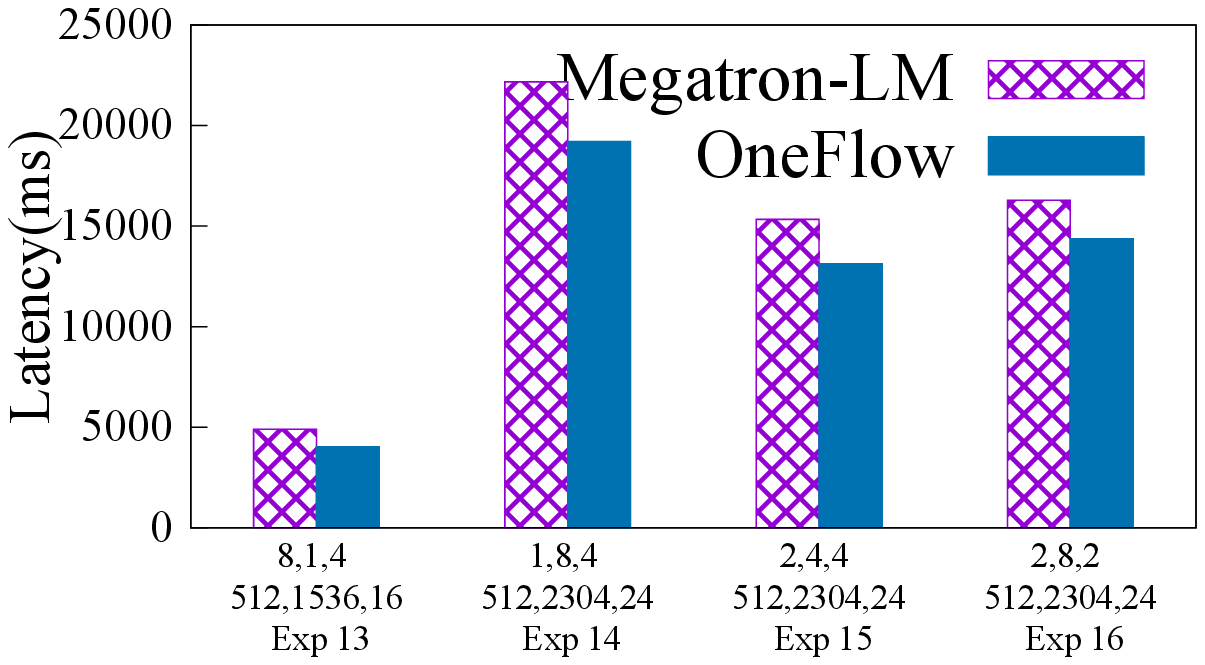}
         \caption{DP, MP combined with PP}
     \end{subfigure}

     \caption{Per-iteration training time for training GPT-2 using various parallelisms: \sys{} vs.~Megatron-LM. The numbers listed for each experiment 
     are respectively data-parallel-size, tensor-model-parallel-size, pipeline-model-parallel-size, global batch size, hidden-size, number-of-layers defined in Megatron-LM.}
     \vspace{-0.3cm}
     \label{figure:megatron_latency_memory}
\end{figure}

\vspace{-0.2cm}
Megatron-LM \cite{mohanmmad-megatron-2020} is a customized library for pre-training large 
models such as GPT-3 based on PyTorch. It 
supports data parallelism, 
model parallelism 
and hybrid parallelism which combines data and model parallelism (amounting to the two-dimensional \textit{SBP} described in Section \ref{section:gshard}). It also implements activation checkpointing and synchronous pipeline with 1F1B pipeline schedule. 
We compare \SysName{} and Megatron-LM for training GPT-2 under representative configurations in Figure \ref{figure:megatron_latency_memory}. The four sub-figures demonstrates the experiment results for pure data parallelism, pure model parallelism, hybrid of data parallelism and model parallelism, a combination of data, model and pipeline parallelism. As a generic framework, \SysName{} implements all features that Megatron-LM supports, such as the activation checkpointing and 1F1B pipeline schedule techniques and align all the hyper-parameters. The physical execution plans of two frameworks are essentially the same. However, \SysName{} performs more kernel fusions than Megatron-LM does. In the result, \SysName{} outperforms Megatron-LM even with a single device. This is the major reason why \SysName{} achieves 
higher training efficiency in distributed cases over the customized library.

\vspace{-0.3cm}
\section{Conclusion and Discussions}
\vspace{-0.2cm}


We propose a new distributed deep learning framework \SysName{} based on the concept of \textit{SBP} and the actor model. \SysName{} overcomes the complexity and efficiency issues of existing frameworks in supporting various parallelisms 
for training large DL models. The compiler uses the concise abstraction of \textit{SBP} for automatically generating an effective execution plan for actors with both spatial and temporal scheduling enabled. The actor model unifies various dependencies as message passing and naturally supports pipelining,  serving a novel mechanism for runtime of distributed DL frameworks. Finally, we show experiment results from a wide range of  challenging tasks on real datasets to demonstrate that the design presented in this paper is more flexible and efficient than the existing ones.

Even though both \SysName{} and Ray~\cite{phillipp-2017-ray} use the concept of the actor, the granularities are different. In Ray, a single actor is used to manage a complete neural network while performing deep learning training. So far, Ray can only act as a plugin to enable data-parallelism to TensorFlow and PyTorch. It does not support model parallelism and pipeline parallelism.


There are still a number of areas that we are actively working on to improve \SysName{}, including:
(1) to enable \SysName{} with elastic scaling~\cite{mai2020kungfu, or2020resource} and fine-grained fault resilience~\cite{wang2021ownership, zaharia2013discretized} besides the naive global checkpointing;
(2) to implement auto placement and auto parallelism by designing a more efficient cost model, thus making it easier to use.


\section*{Acknowledgements}

We thank the anonymous reviewers of OSDI 2021, SOSP 2021 and MLSys 2022 for their helpful comments on the paper. Developing a deep learning framework such as \SysName{} involves a large amount of engineering efforts. We gratefully acknowledge contributions from our colleagues within OneFlow Inc. and Zhejiang Lab., and from the users of \SysName{}. In particular, Wenxiao Zhang, Xiaoyu Zhang, Binbin Han, Jianhao Zhang, Houjiang Chen, Luyang Zhao, Yu Ouyang, Zekang Zheng, Xuan Xie, Yinggang Wang, Yipeng Li, Fengwei Liu, Shijie Wang, Xiaoyu Xu, Depeng Liang, Mingyang Liu, Shiyuan Shangguan, Jing Qiao, Chong Niu, Wei Zhang, Xuefei Jiang contribute a lot of code to \SysName{}.

\bibliography{of}

\begin{thebibliography}{50}
\providecommand{\natexlab}[1]{#1}
\providecommand{\url}[1]{\texttt{#1}}
\expandafter\ifx\csname urlstyle\endcsname\relax
  \providecommand{\doi}[1]{doi: #1}\else
  \providecommand{\doi}{doi: \begingroup \urlstyle{rm}\Url}\fi

\bibitem[dee(2021)]{deepspeed}
{Microsoft DeepSpeed}.
\newblock \url{https://github.com/microsoft/DeepSpeed}, 2021.

\bibitem[ins(2021)]{insightface}
{InsightFace Project}.
\newblock \url{https://github.com/deepinsight/insightface}, 2021.

\bibitem[jea(2021)]{jeaugey2017nccl}
{NVIDIA NCCL}.
\newblock \url{https://developer.nvidia.com/nccl}, 2021.

\bibitem[nvi(2021)]{nvidia-dali}
{NVIDIA Data Loading Library (DALI)}.
\newblock \url{https://developer.nvidia.com/DALI}, 2021.

\bibitem[Abadi et~al.(2016)Abadi, Barham, Chen, Chen, Davis, Dean, Devin,
  Ghemawat, Irving, Isard, Kudlur, Levenberg, Monga, Moore, Murray, Steiner,
  Tucker, Vasudevan, Warden, Wicke, Yu, and Zheng]{abadi-2016-tensorflow}
Abadi, M., Barham, P., Chen, J., Chen, Z., Davis, A., Dean, J., Devin, M.,
  Ghemawat, S., Irving, G., Isard, M., Kudlur, M., Levenberg, J., Monga, R.,
  Moore, S., Murray, D.~G., Steiner, B., Tucker, P., Vasudevan, V., Warden, P.,
  Wicke, M., Yu, Y., and Zheng, X.
\newblock {TensorFlow: A System for Large-scale Machine Learning}.
\newblock In \emph{Proceedings of the 12th USENIX Conference on Operating
  Systems Design and Implementation}, 2016.

\bibitem[Ben-Nun \& Hoefler(2019)Ben-Nun and Hoefler]{tal-ddl-2019}
Ben-Nun, T. and Hoefler, T.
\newblock {Demystifying Parallel and Distributed Deep Learning: An In-Depth
  Concurrency Analysis}.
\newblock \emph{ACM Computing Surveys}, 2019.

\bibitem[Brown et~al.(2020)Brown, Mann, Ryder, Subbiah, Kaplan, Dhariwal,
  Neelakantan, Shyam, Sastry, Askell, Agarwal, Herbert-Voss, Krueger, Henighan,
  Child, Ramesh, Ziegler, Wu, Winter, Hesse, Chen, Sigler, Litwin, Gray, Chess,
  Clark, Berner, McCandlish, Radford, Sutskever, and Amodei]{brown-gpt-2020}
Brown, T., Mann, B., Ryder, N., Subbiah, M., Kaplan, J.~D., Dhariwal, P.,
  Neelakantan, A., Shyam, P., Sastry, G., Askell, A., Agarwal, S.,
  Herbert-Voss, A., Krueger, G., Henighan, T., Child, R., Ramesh, A., Ziegler,
  D., Wu, J., Winter, C., Hesse, C., Chen, M., Sigler, E., Litwin, M., Gray,
  S., Chess, B., Clark, J., Berner, C., McCandlish, S., Radford, A., Sutskever,
  I., and Amodei, D.
\newblock {Language Models are Few-Shot Learners}.
\newblock In \emph{Proceedings of Advances in Neural Information Processing
  Systems}, 2020.

\bibitem[Chen et~al.(2016{\natexlab{a}})Chen, Monga, Bengio, and
  J{\'{o}}zefowicz]{jiamin-2016-corr}
Chen, J., Monga, R., Bengio, S., and J{\'{o}}zefowicz, R.
\newblock {Revisiting Distributed Synchronous {SGD}}.
\newblock \emph{arXiv preprint arXiv:1604.00981}, 2016{\natexlab{a}}.

\bibitem[Chen et~al.(2018)Chen, Liu, Gao, and Han]{chen2018mobilefacenets}
Chen, S., Liu, Y., Gao, X., and Han, Z.
\newblock {Mobilefacenets: Efficient CNNs for Accurate Real-time Face
  Verification on Mobile Devices}.
\newblock In \emph{Proceedings of Chinese Conference on Biometric Recognition},
  2018.

\bibitem[Chen et~al.(2015)Chen, Li, Li, Lin, Wang, Wang, Xiao, Xu, Zhang, and
  Zhang]{chen-2015-mxnet}
Chen, T., Li, M., Li, Y., Lin, M., Wang, N., Wang, M., Xiao, T., Xu, B., Zhang,
  C., and Zhang, Z.
\newblock {MXNet: A Flexible and Efficient Machine Learning Library for
  Heterogeneous Distributed System}.
\newblock In \emph{Proceedings of LearningSys}, 2015.

\bibitem[Chen et~al.(2016{\natexlab{b}})Chen, Xu, Zhang, and
  Guestrin]{chen2016training}
Chen, T., Xu, B., Zhang, C., and Guestrin, C.
\newblock {Training Deep Nets with Sublinear Memory Cost}.
\newblock \emph{arXiv preprint arXiv:1604.06174}, 2016{\natexlab{b}}.

\bibitem[Cheng et~al.(2016)Cheng, Koc, Harmsen, Shaked, Chandra, Aradhye,
  Anderson, Corrado, Chai, Ispir, Anil, Haque, Hong, Jain, Liu, and
  Shah]{cheng2016wide}
Cheng, H.-T., Koc, L., Harmsen, J., Shaked, T., Chandra, T., Aradhye, H.,
  Anderson, G., Corrado, G., Chai, W., Ispir, M., Anil, R., Haque, Z., Hong,
  L., Jain, V., Liu, X., and Shah, H.
\newblock {Wide \& Deep Learning for Recommender Systems}.
\newblock In \emph{Proceedings of the 1st Workshop on Deep Learning for
  Recommender Systems}, 2016.

\bibitem[Devlin et~al.(2019)Devlin, Chang, Lee, and
  Toutanova]{devlin-etal-2019-bert}
Devlin, J., Chang, M.-W., Lee, K., and Toutanova, K.
\newblock {{BERT}: Pre-training of Deep Bidirectional Transformers for Language
  Understanding}.
\newblock In \emph{Proceedings of the North American Chapter of the Association
  for Computational Linguistics: Human Language Technologies}, 2019.

\bibitem[fairscale()]{fairscale}
fairscale.
\newblock {Facebook Fairscale project}.
\newblock \url{https://github.com/facebookresearch/fairscale}.

\bibitem[Fedus et~al.(2021)Fedus, Zoph, and Shazeer]{fedus-switch-2021}
Fedus, W., Zoph, B., and Shazeer, N.
\newblock {Switch Transformers: Scaling to Trillion Parameter Models with
  Simple and Efficient Sparsity}.
\newblock \emph{arXiv preprint arXiv:2101.03961}, 2021.

\bibitem[Goyal et~al.(2017)Goyal, Doll¨¢r, Girshick, Noordhuis, Wesolowski,
  Kyrola, Tulloch, Jia, and He]{priya-2017-large}
Goyal, P., Doll¨¢r, P., Girshick, R., Noordhuis, P., Wesolowski, L., Kyrola,
  A., Tulloch, A., Jia, Y., and He, K.
\newblock {Accurate, Large Minibatch SGD: Training ImageNet in 1 Hour}.
\newblock \emph{arXiv preprint arXiv:1706.02677}, 2017.

\bibitem[Hashemi et~al.(2019)Hashemi, Jyothi, and Campbell]{hashemi19tictac}
Hashemi, S.~H., Jyothi, S.~A., and Campbell, R.~H.
\newblock {TicTac: Accelerating Distributed Deep Learning with Communication
  Scheduling}.
\newblock In \emph{Proceedings of Machine Learning and Systems}, 2019.

\bibitem[{He} et~al.(2016){He}, {Zhang}, {Ren}, and {Sun}]{he2016deep}
{He}, K., {Zhang}, X., {Ren}, S., and {Sun}, J.
\newblock {Deep Residual Learning for Image Recognition}.
\newblock In \emph{Proceedings of the IEEE Conference on Computer Vision and
  Pattern Recognition}, 2016.

\bibitem[Hewitt et~al.(1973)Hewitt, Bishop, and Steiger]{hewitt-ijcai-1973}
Hewitt, C., Bishop, P., and Steiger, R.
\newblock {A Universal Modular ACTOR Formalism for Artificial Intelligence}.
\newblock In \emph{Proceedings of the 3rd International Joint Conference on
  Artificial Intelligence}, 1973.

\bibitem[Huang et~al.(2019)Huang, Cheng, Bapna, Firat, Chen, Chen, Lee, Ngiam,
  Le, Wu, and Chen]{huang-2019-nips}
Huang, Y., Cheng, Y., Bapna, A., Firat, O., Chen, D., Chen, M., Lee, H., Ngiam,
  J., Le, Q.~V., Wu, Y., and Chen, z.
\newblock {GPipe: Efficient Training of Giant Neural Networks using Pipeline
  Parallelism}.
\newblock In \emph{Proceedings of Advances in Neural Information Processing
  Systems}, 2019.

\bibitem[Jia et~al.(2018)Jia, Lin, Qi, and Aiken]{jia-2018-icml}
Jia, Z., Lin, S., Qi, C.~R., and Aiken, A.
\newblock {Exploring Hidden Dimensions in Parallelizing Convolutional Neural
  Networks}.
\newblock In \emph{Proceedings of the International Conference on Machine
  Learning}, 2018.

\bibitem[Jia et~al.(2019)Jia, Zaharia, and Aiken]{jia-2019-sysml}
Jia, Z., Zaharia, M., and Aiken, A.
\newblock {Beyond Data and Model Parallelism for Deep Neural Networks}.
\newblock In \emph{Proceedings of Machine Learning and Systems}, 2019.

\bibitem[Jiang et~al.(2020)Jiang, Zhu, Lan, Yi, Cui, and Guo]{jiang-osdi-2020}
Jiang, Y., Zhu, Y., Lan, C., Yi, B., Cui, Y., and Guo, C.
\newblock {A Unified Architecture for Accelerating Distributed {DNN} Training
  in Heterogeneous GPU/CPU Clusters}.
\newblock In \emph{Proceedings of the 14th USENIX Symposium on Operating
  Systems Design and Implementation}, 2020.

\bibitem[Kaplan et~al.(2020)Kaplan, McCandlish, Henighan, Brown, Chess, Child,
  Gray, Radford, Wu, and Amodei]{kaplan-scaling-2020}
Kaplan, J., McCandlish, S., Henighan, T., Brown, T.~B., Chess, B., Child, R.,
  Gray, S., Radford, A., Wu, J., and Amodei, D.
\newblock {Scaling Laws for Neural Language Models}.
\newblock \emph{arXiv preprint arXiv:2001.08361}, 2020.

\bibitem[Kingma \& Ba(2015)Kingma and Ba]{kingma2014adam}
Kingma, D.~P. and Ba, J.
\newblock {Adam: A Method for Stochastic Optimization}.
\newblock In \emph{Proceedings of International Conference on Learning
  Representations}, 2015.

\bibitem[Kumar et~al.(2020)Kumar, Bradbury, Young, Wang, Levskaya, Hechtman,
  Chen, Lee, Deveci, Kumar, et~al.]{kumar2020exploring}
Kumar, S., Bradbury, J., Young, C., Wang, Y.~E., Levskaya, A., Hechtman, B.,
  Chen, D., Lee, H., Deveci, M., Kumar, N., et~al.
\newblock {Exploring the Limits of Concurrency in ML Training on Google TPUs}.
\newblock \emph{arXiv preprint arXiv:2011.03641}, 2020.

\bibitem[Kung et~al.(1994)Kung, Blackwell, and Chapman]{kung-credit-1994}
Kung, H.~T., Blackwell, T., and Chapman, A.
\newblock Credit-based flow control for atm networks: Credit update protocol,
  adaptive credit allocation and statistical multiplexing.
\newblock \emph{SIGCOMM Comput. Commun. Rev.}, 24\penalty0 (4):\penalty0
  101–114, October 1994.

\bibitem[Lepikhin et~al.(2020)Lepikhin, Lee, Xu, Chen, Firat, Huang, Krikun,
  Shazeer, and Chen]{dmitry-2020-gshard}
Lepikhin, D., Lee, H., Xu, Y., Chen, D., Firat, O., Huang, Y., Krikun, M.,
  Shazeer, N., and Chen, Z.
\newblock {GShard: Scaling Giant Models with Conditional Computation and
  Automatic Sharding}.
\newblock In \emph{Proceedings of International Conference on Learning
  Representations}, 2020.

\bibitem[Li et~al.(2014)Li, Andersen, Park, Smola, Ahmed, Josifovski, Long,
  Shekita, and Su]{li14ps}
Li, M., Andersen, D.~G., Park, J.~W., Smola, A.~J., Ahmed, A., Josifovski, V.,
  Long, J., Shekita, E.~J., and Su, B.-Y.
\newblock {Scaling Distributed Machine Learning with the Parameter Server}.
\newblock In \emph{Proceedings of the 11th USENIX Symposium on Operating
  Systems Design and Implementation}, 2014.

\bibitem[Mai et~al.(2020)Mai, Li, Wagenl{\"a}nder, Fertakis, Brabete, and
  Pietzuch]{mai2020kungfu}
Mai, L., Li, G., Wagenl{\"a}nder, M., Fertakis, K., Brabete, A.-O., and
  Pietzuch, P.
\newblock {KungFu: Making Training in Distributed Machine Learning Adaptive}.
\newblock In \emph{Proceedings of the 14th USENIX Symposium on Operating
  Systems Design and Implementation}, 2020.

\bibitem[Mattson et~al.(2020)Mattson, Cheng, Diamos, Coleman, Micikevicius,
  Patterson, Tang, Wei, Bailis, Bittorf, Brooks, Chen, Dutta, Gupta, Hazelwood,
  Hock, Huang, Kang, Kanter, Kumar, Liao, Narayanan, Oguntebi, Pekhimenko,
  Pentecost, Janapa~Reddi, Robie, St~John, Wu, Xu, Young, and
  Zaharia]{mlperf-training}
Mattson, P., Cheng, C., Diamos, G., Coleman, C., Micikevicius, P., Patterson,
  D., Tang, H., Wei, G.-Y., Bailis, P., Bittorf, V., Brooks, D., Chen, D.,
  Dutta, D., Gupta, U., Hazelwood, K., Hock, A., Huang, X., Kang, D., Kanter,
  D., Kumar, N., Liao, J., Narayanan, D., Oguntebi, T., Pekhimenko, G.,
  Pentecost, L., Janapa~Reddi, V., Robie, T., St~John, T., Wu, C.-J., Xu, L.,
  Young, C., and Zaharia, M.
\newblock {MLPerf Training Benchmark}.
\newblock In \emph{Proceedings of Machine Learning and Systems}, 2020.

\bibitem[Micikevicius et~al.(2018)Micikevicius, Narang, Alben, Diamos, Elsen,
  Garcia, Ginsburg, Houston, Kuchaiev, Venkatesh, and
  Wu]{micikevicius2018mixed}
Micikevicius, P., Narang, S., Alben, J., Diamos, G., Elsen, E., Garcia, D.,
  Ginsburg, B., Houston, M., Kuchaiev, O., Venkatesh, G., and Wu, H.
\newblock {Mixed Precision Training}.
\newblock In \emph{Proceedings of International Conference on Learning
  Representations}, 2018.

\bibitem[Moritz et~al.(2018)Moritz, Nishihara, Wang, Tumanov, Liaw, Liang,
  Paul, Jordan, and Stoica]{phillipp-2017-ray}
Moritz, P., Nishihara, R., Wang, S., Tumanov, A., Liaw, R., Liang, E., Paul,
  W., Jordan, M.~I., and Stoica, I.
\newblock {Ray: A Distributed Framework for Emerging AI Applications}.
\newblock In \emph{Proceedings of the 13th USENIX Symposium on Operating
  Systems Design and Implementation}, 2018.

\bibitem[Narayanan et~al.(2019)Narayanan, Harlap, Phanishayee, Seshadri,
  Devanur, Ganger, Gibbons, and Zaharia]{narayanan-2019-sosp}
Narayanan, D., Harlap, A., Phanishayee, A., Seshadri, V., Devanur, N.~R.,
  Ganger, G.~R., Gibbons, P.~B., and Zaharia, M.
\newblock {PipeDream: Generalized Pipeline Parallelism for DNN Training}.
\newblock In \emph{Proceedings of the 27th ACM Symposium on Operating Systems
  Principles}, 2019.

\bibitem[Narayanan et~al.(2021)Narayanan, Shoeybi, Casper, LeGresley, Patwary,
  Korthikanti, Vainbrand, Kashinkunti, Bernauer, Catanzaro, Phanishayee, and
  Zaharia]{narayanan-megatron-2021}
Narayanan, D., Shoeybi, M., Casper, J., LeGresley, P., Patwary, M.,
  Korthikanti, V., Vainbrand, D., Kashinkunti, P., Bernauer, J., Catanzaro, B.,
  Phanishayee, A., and Zaharia, M.
\newblock {Efficient Large-Scale Language Model Training on GPU Clusters}.
\newblock \emph{arXiv preprint arXiv:2104.04473}, 2021.

\bibitem[Oldridge et~al.(2020)Oldridge, Perez, Frederickson, Koumchatzky, Lee,
  Wang, Wu, Yu, Zamora, Yılmaz, Gunny, Nguyen, and Lee]{Oldridge2020MerlinAG}
Oldridge, E., Perez, J., Frederickson, B., Koumchatzky, N., Lee, M., Wang,
  Z.-H., Wu, L., Yu, F., Zamora, R., Yılmaz, O., Gunny, A.~M., Nguyen, V.~P.,
  and Lee, S.
\newblock {Merlin: A GPU Accelerated Recommendation Framework}.
\newblock 2020.

\bibitem[Or et~al.(2020)Or, Zhang, and Freedman]{or2020resource}
Or, A., Zhang, H., and Freedman, M.
\newblock {Resource Elasticity in Distributed Deep Learning}.
\newblock 2020.

\bibitem[Pal et~al.(2019)Pal, Ebrahimi, Zulfiqar, Fu, Zhang, Migacz, Nellans,
  and Gupta]{pal2019optimizing}
Pal, S., Ebrahimi, E., Zulfiqar, A., Fu, Y., Zhang, V., Migacz, S., Nellans,
  D., and Gupta, P.
\newblock {Optimizing Multi-GPU Parallelization Strategies for Deep Learning
  Training}.
\newblock \emph{IEEE Micro}, 2019.

\bibitem[Paszke et~al.(2019)Paszke, Gross, Massa, Lerer, Bradbury, Chanan,
  Killeen, Lin, Gimelshein, Antiga, Desmaison, Kopf, Yang, DeVito, Raison,
  Tejani, Chilamkurthy, Steiner, Fang, Bai, and Chintala]{paszke-2019-pytorch}
Paszke, A., Gross, S., Massa, F., Lerer, A., Bradbury, J., Chanan, G., Killeen,
  T., Lin, Z., Gimelshein, N., Antiga, L., Desmaison, A., Kopf, A., Yang, E.,
  DeVito, Z., Raison, M., Tejani, A., Chilamkurthy, S., Steiner, B., Fang, L.,
  Bai, J., and Chintala, S.
\newblock {PyTorch: An Imperative Style, High-Performance Deep Learning
  Library}.
\newblock In \emph{Proceedings of Advances in Neural Information Processing
  Systems}, 2019.

\bibitem[Peng et~al.(2019)Peng, Zhu, Chen, Bao, Yi, Lan, Wu, and
  Guo]{peng-2019-sosp}
Peng, Y., Zhu, Y., Chen, Y., Bao, Y., Yi, B., Lan, C., Wu, C., and Guo, C.
\newblock {A Generic Communication Scheduler for Distributed DNN Training
  Acceleration}.
\newblock In \emph{Proceedings of the 27th ACM Symposium on Operating Systems
  Principles}, 2019.

\bibitem[Rajbhandari et~al.(2020)Rajbhandari, Rasley, Ruwase, and
  He]{rajbhandari-zero-2020}
Rajbhandari, S., Rasley, J., Ruwase, O., and He, Y.
\newblock {ZeRO: Memory Optimizations Toward Training Trillion Parameter
  Models}.
\newblock In \emph{Proceedings of International Conference for High Performance
  Computing, Networking, Storage and Analysis}, 2020.

\bibitem[Rajbhandari et~al.(2021)Rajbhandari, Ruwase, Rasley, Smith, and
  He]{rajbhandari-zeroinfinity-2021}
Rajbhandari, S., Ruwase, O., Rasley, J., Smith, S., and He, Y.
\newblock {ZeRO-Infinity: Breaking the GPU Memory Wall for Extreme Scale Deep
  Learning}.
\newblock \emph{arXiv preprint arXiv:2104.07857}, 2021.

\bibitem[Sergeev \& Balso(2018)Sergeev and Balso]{alexander-2018-horovod}
Sergeev, A. and Balso, M.~D.
\newblock {Horovod: Fast and Easy Distributed Deep Learning in TensorFlow}.
\newblock \emph{arXiv preprint arXiv:1802.05799}, 2018.

\bibitem[Shazeer et~al.(2018)Shazeer, Cheng, Parmar, Tran, Vaswani,
  Koanantakool, Hawkins, Lee, Hong, Young, et~al.]{shazeer-2018-nips}
Shazeer, N., Cheng, Y., Parmar, N., Tran, D., Vaswani, A., Koanantakool, P.,
  Hawkins, P., Lee, H., Hong, M., Young, C., et~al.
\newblock {Mesh-tensorflow: Deep learning for supercomputers}.
\newblock In \emph{Proceedings of Advances in Neural Information Processing
  Systems}, 2018.

\bibitem[Shoeybi et~al.(2020)Shoeybi, Patwary, Puri, LeGresley, Casper, and
  Catanzaro]{mohanmmad-megatron-2020}
Shoeybi, M., Patwary, M., Puri, R., LeGresley, P., Casper, J., and Catanzaro,
  B.
\newblock {Megatron-LM: Training Multi-Billion Parameter Language Models Using
  Model Parallelism}.
\newblock \emph{arXiv preprint arXiv:1909.08053}, 2020.

\bibitem[Wang et~al.(2019)Wang, Huang, and Li]{wang-2019-eurosys}
Wang, M., Huang, C.-c., and Li, J.
\newblock {Supporting Very Large Models Using Automatic Dataflow Graph
  Partitioning}.
\newblock In \emph{Proceedings of the Fourteenth EuroSys Conference}, 2019.

\bibitem[Wang et~al.(2021)Wang, Liang, Oakes, Hindman, Luan, Cheng, and
  Stoica]{wang2021ownership}
Wang, S., Liang, E., Oakes, E., Hindman, B., Luan, F.~S., Cheng, A., and
  Stoica, I.
\newblock {Ownership: A Distributed Futures System for Fine-Grained Tasks}.
\newblock In \emph{Proceedings of the 18th USENIX Symposium on Networked
  Systems Design and Implementation}, 2021.

\bibitem[Wu et~al.(2018)Wu, Xu, Li, and Xiong]{wu2018stanza}
Wu, X., Xu, H., Li, B., and Xiong, Y.
\newblock {Stanza: Distributed Deep Learning with Small Communication
  Footprint}.
\newblock \emph{arXiv preprint arXiv:1812.10624}, 2018.

\bibitem[Xu et~al.(2021)Xu, Li, Gong, and You]{xu2021efficient}
Xu, Q., Li, S., Gong, C., and You, Y.
\newblock An efficient 2d method for training super-large deep learning models,
  2021.

\bibitem[Zaharia et~al.(2013)Zaharia, Das, Li, Hunter, Shenker, and
  Stoica]{zaharia2013discretized}
Zaharia, M., Das, T., Li, H., Hunter, T., Shenker, S., and Stoica, I.
\newblock Discretized streams: Fault-tolerant streaming computation at scale.
\newblock In \emph{Proceedings of the twenty-fourth ACM Symposium on Operating
  Systems Principles}, 2013.

\end{thebibliography}
\bibliographystyle{mlsys2022}

%


\end{document}